\documentclass[12pt,preprint]{aastex} 

\newcommand{\be}{\begin{equation}}
\newcommand{\ee}{\end{equation}}
\newcommand{\bea}{\begin{eqnarray}}
\newcommand{\eea}{\end{eqnarray}}
\newcommand{\ud}{\mathrm{d}}

\shorttitle{Galaxy Orbits for Galaxy Clusters} 
\shortauthors{Hwang \& Lee}

\begin{document}

\title{Galaxy Orbits for Galaxy Clusters in Sloan Digital Sky
Survey and 2dF Galaxy Redshift Survey}

\author{Ho Seong Hwang\altaffilmark{1,2} and Myung Gyoon Lee\altaffilmark{1}}
\altaffiltext{1}{Astronomy Program, Department of Physics and Astronomy,
Seoul National University, 56-1 Sillim 9-dong, Gwanak-gu, Seoul 151-742, Korea}
\altaffiltext{2}{School of Physics, Korea Institute for Advanced Study, Seoul 130-722, Korea}
\email{hshwang@kias.re.kr, mglee@astro.snu.ac.kr}

\begin{abstract}
We present the results of a study for galaxy orbits in galaxy
  clusters using a spectroscopic sample of galaxies in Sloan Digital
  Sky Survey (SDSS) and 2dF Galaxy Redshift Survey (2dFGRS).
We have determined the member galaxies of Abell clusters covered
  by these surveys using the galaxies' redshift and positional data.
We have selected 10 clusters using three criteria: 
  the number of member galaxies is greater than or equal to 40, 
  the spatial coverage is complete, and
  X-ray mass profile is available in the literature.
We derive the radial profile of the galaxy number density and
  velocity dispersion using all, early-type, and late-type galaxies
  for each cluster.
We have investigated the galaxy orbits for our sample clusters with
  constant and variable velocity anisotropies over the clustercentric
  distance using Jeans equation. 
Using all member galaxies,
  the galaxy orbits are found to be isotropic within the uncertainty for most of sample clusters,
  although it is difficult to conclude strongly for some clusters
  due the large errors and the variation as a function of the clustercentric distance
  in the calculated velocity anisotropies.
We investigated the orbital difference between early-type and late-type galaxies 
  for four sample clusters,
  and found no significant difference between them.
\end{abstract}



\section{Introduction}

The mass estimate of a galaxy cluster, the largest gravitationally bounded system in the universe,
was given for the first time by \citet{zwi33},
which showed an indication of the existence of dark matter.
In general, the cluster mass
ranges from $10^{13}M_\odot$ for small groups to $10^{15} M_\odot$ for rich clusters,
which is in the form of galaxies ($1-2\%$ of the total mass, e.g., \citealt{lin03}),
hot X-ray emitting gas ($5-15\%$, e.g., \citealt{vik06}),
dark matter ($80-90\%$) that affects the galaxy cluster through the gravitation only.

To determine the amount and distribution of underlying dark matter in clusters, 
  Jeans analysis using the positional and velocity data of cluster galaxies
  has been usually adopted among several mass estimation methods \citep{kbm04,biv06}. 
However, previous studies had to assume the galaxy orbit in prior to derive
  the mass profile of galaxy clusters, due to the `mass-orbit'
  degeneracy in the velocity dispersion profile (VDP) (e.g., \citealt{mer87}). 
This degeneracy can be broken 
 by an orbit analysis with an independent mass determination 
   based on X-ray or lensing studies,
 or by an analysis of higher order moments of velocity distribution :
  velocity dispersion and kurtosis (e.g., \citealt{lm03,lok06}),
  Gauss-Hermite moments (e.g., \citealt{van00,kbm04}).

The analysis of galaxy orbits for clusters was made for the
Coma cluster for the first time by \citet{kg82} who used about 300 galaxy velocities.
They showed that the galaxy orbits in Coma can not be primarily
radial, and a significant fraction of kinetic energy must be in
tangential direction even at large radii. 
With an aid of large redshift surveys for cluster galaxies such as 
  Canadian Network for Observational Cosmology (CNOC; \citealt{yee96}), 
  ESO Nearby Abell Cluster Survey (ENACS; \citealt{kat96}), and
  Cluster and Infall Region Nearby Survey (CAIRNS; \citealt{rines03}), 
  extensive analysis for galaxy orbits in clusters has been performed.
\citet{car97a,car97b} analyzed 14 CNOC clusters at $z=0.17-0.54$,
and found that the galaxy orbits are isotropic or modest radial.
\citet{van00}, using the same CNOC data, concluded that their
best-fit model is close to isotropic. \citet{rines03} secured
15,000 galaxies in CAIRNS data, and concluded that the galaxy
orbits in their ensemble cluster are also consistent with being isotropic.

On the other hand, the difference in galaxy orbits and in galaxy
  velocity dispersions among galaxy types provide important clues to
  understand the formation history of the cluster \citep{bk04,goto05}. 
\citet{biv97} studied kinematic difference
  between emission-line galaxies (ELGs) and non-ELGs using ENACS data, 
  and found that the VDP of ELGs is consistent with being radial, 
  while that of non-ELGs is not. 
\citet{abm98}, based on simple modelling of observed VDP for about 2000 galaxies 
  in 40 regular clusters, reported that the orbits of ellipticals are
  mostly tangential in the cluster core and are nearly isotropic outside, 
  while those of spirals are predominantly radial.
\citet{mah99} analyzed a sample of 20 galaxy groups from Center
  for Astrophysics redshift survey. 
They found that star-forming galaxies or ELGs have moderately radial orbits, 
  while old or absorption-line galaxies isotropic orbits within the errors. 
Recently, \citet{bk04}, using the ENACS data, 
  studied the galaxy orbits of all galaxy classes 
  (the brightest ellipticals, other ellipticals together with S0,
  early-type spirals, late-type spirals, and irregulars) for the first time. 
They reported that 
  the brightest ellipticals do not yield equilibrium solution, 
  other ellipticals together with S0 have isotropic orbits as given in \citet{kbm04},
  and early spirals are consistent with isotropic orbits, 
  while late spirals prefer radial orbits to isotropic orbits. 
In contrast, \citet{rd98}, using nearby ($z<0.055$) Abell clusters, 
  concluded that the orbits of elliptical galaxies in clusters are close to radial, 
  while those of spirals more tangential or isotropic. 
For nine intermediate ($z\sim0.1-0.4$) CNOC cluster, 
  \citet{rds00} obtained similar results to those of \citet{rd98}: 
  bulge-dominated galaxies have more eccentric orbits than disk-dominated galaxies do.
The cause for the different results of Ram{\'{\i}}rez et al. compared with other studies is discussed
  in detail in \citet{van00} and \citet{biv02}.

Above results of galaxy orbits are based on a composite
clusters prepared by combining data for disparate clusters that might have different formation histories.
Although it is helpful to make the composite cluster in order to overcome problems of
1) a limited number of measured galaxy velocities per cluster and 2) an application of spherical Jeans
equation to an asymmetric cluster, the composite cluster might be significantly different from any real cluster.
In addition, previous studies were not based on the independent measurements
of cluster mass profiles, but based on the optical galaxy data.
These problems can be overcome. \citet{nk96} demonstrated the galaxy orbits can be constrained
using an independently determined mass profile. For A2218, they
used mass estimates derived by combining strong and weak lensing effects,
and found evidence for an anisotropic core.
Later, \citet{ben06} extended the study of galaxy orbits to five clusters
using mass estimates from X-ray data.
They reported that the galaxies in five clusters have diverse orbits,
and the orbital profiles in observed and simulated clusters appear to be different:
simulated clusters show preferentially tangential orbits.
However, they did not divide the galaxy sample into different types
to investigate the orbital properties of different galaxy types.

Recently, galaxy redshift surveys such as the Sloan Digital
Sky Survey (SDSS; \citealt{york00}) and the 2dF Galaxy Redshift Survey
(2dFGRS; \citealt{col01}) have provided redshift data for large samples of
galaxies. In addition, X-ray mass profiles for numerous clusters
become available \citep{rb02,san03,dem03,bm06,vik06,vf06}.
By identifying member galaxies in clusters using these redshift and positional data
  and adopting X-ray mass profiles,
  we can investigate the galaxy orbits in clusters for a large sample of galaxy clusters
  without making the composite cluster.
We can also investigate the orbital properties for different galaxy types.

In this paper, we present the results of a study for the orbits of galaxies in galaxy clusters,
using a spectroscopic sample of galaxies in SDSS and 2dFGRS and X-ray mass profiles in the literature.
Section \ref{data} describes the sample of galaxies and clusters used in this study.
Radial profiles of galaxy number density and velocity dispersion are derived in \S \ref{property}.
Analysis of the galaxy orbits and global kinematics are given in \S \ref{result} and \S \ref{global}, respectively.
Discussion and summary are given in \S \ref{discuss} and \S \ref{sum}, respectively.
Throughout, we adopt
cosmological parameters $h=0.7$, $\Omega_{\Lambda}=0.7$, and
$\Omega_{M}=0.3$.

\section{Data}\label{data}
\subsection {Galaxy Sample}\label{gal}
We used data from the spectroscopic sample of galaxies in the
Legacy survey of SDSS Sixth Data Release\footnote{Access to Data
  Release 6 can be found on the SDSS Web site
  (http://www.sdss.org/dr6).} \citep{str07},
  and in the 2dFGRS Final Data Release \citep{col01}. 
SDSS is one of the largest imaging and spectroscopic survey, which covers more
than a quarter of the sky \citep{york00} using a dedicated,
2.5-meter telescope \citep{gunn06} at Apache Point Observatory,
New Mexico. The telescope is equipped with
  a wide-field drift-scanning mosaic CCD camera
   that can image 1.5 deg$^2$ of sky at a time for imaging survey \citep{gunn98}, and
  a pair of fiber-fed spectrographs 
   that can measure spectra more than 600 objects in a single observation
   with 3$^\circ$ field of view \citep{uom99,cas01,bla03}.
Extensive description of SDSS data products is given by
\citet{york00} and \citet{sto02}, and we only give a brief summary
of the SDSS data. The imaging survey is carried out with five
broad bands ({\it ugriz})
  centered at 3551, 4686, 6166, 7480, and 8932 \AA ~\citep{fuk96}.
  The magnitude limits of five bands for point sources
  are 22.0, 22.2, 22.2, 21.3, and 20.5, respectively.
The calibration of data products from imaging survey
 is given by several authors: data reduction and photometry \citep{lup02},
photometric calibration \citep{hogg01,smi02,ive04,tuc06}, and
astrometric calibration \citep{pier03}. The spectroscopic survey
obtains spectra covering 3800$-$9200 \AA, with a wavelength
resolution, $\lambda /\delta \lambda\sim$ 1800. 
There are three kinds of spectroscopic targets: main galaxy sample,
 luminous red galaxies (LRG), and quasar.
The main galaxy sample consists of the galaxies
 with $r$-band Petrosian magnitude $r_{\rm pet}\le17.77$, which are corrected
 for Galactic foreground reddening using \citet{sfd98}.
 The galaxies with $r$-band Petrosian half-light surface brightness $\mu_{50}>24.5$ mag arcsec$^{-2}$
 are rejected; the number ratio of rejected galaxies to all galaxies is only $\sim1\%$.
In addition, the galaxies with 3$\arcsec$ fiber magnitudes
brighter than
 15 magnitude in $g$ or $r$, or brighter than 14.5 magnitude in $i$ are rejected
 to avoid the problems of saturation and cross-talk in the spectrographs.
 A detailed description for main galaxy sample is presented in \citet{str02}.

We used the main galaxy sample of 528,000 galaxies with measured
velocities for this study. The median redshift for this
spectroscopic sample is 0.11. The uncertainties of the redshift
measurements are $\sim$30 km s$^{-1}$. A redshift confidence
parameter (zConf) is assigned from 0 to 1 in the SDSS catalogs. We
used only those galaxies with zConf $\ge$ 0.65.

2dFGRS is a spectroscopic survey for nearly 246,000 galaxies
selected in the photographic $b_{\rm J}$ band from the APM galaxy
catalog \citep{col01}. The survey uses the Two-degree Field (2dF)
multifibre spectrograph on the Anglo-Australian Telescope, which
measure spectra about 400 objects
 in a single observation with 2$^\circ$ field of view \citep{lew02}.
The wavelength coverage of the spectra is 3600$-$8000 \AA~
  with a resolution of 9 \AA.
The spectroscopic targets are the galaxies with extinction
corrected magnitude $b_{\rm J}<19.45$. The survey coverage
consists of two delination stripes and 100 random fields, which
cover in total over 2000 deg$^2$. The median redshift for the sample of
galaxies is similar to that of SDSS.
A redshift quality parameter Q is assigned in the range 1$-$5. The
redshift measurements with Q$\ge$3 are 98.4\% reliable, and have
an overall rms uncertainty of $\sim$85 km s$^{-1}$. We used only
the galaxies with Q$\ge$3 in this study.

Since some galaxies are covered in both SDSS and 2dFGRS, we
matched the galaxies found in SDSS with those in 2dFGRS to make a
master catalog. The mean difference in radial velocity $\Delta v$
($=v_{\rm SDSS}-v_{\rm 2dFGRS}$) between the SDSS and 2dFGRS
measurements for the 29,200 matched galaxies is estimated to be $\sim13$ km
s$^{-1}$. We corrected the SDSS velocities by this mean
difference, and used the average value of the velocities measured
in SDSS and in 2dFGRS for further analysis.

\begin{figure}
\includegraphics [width=135mm]{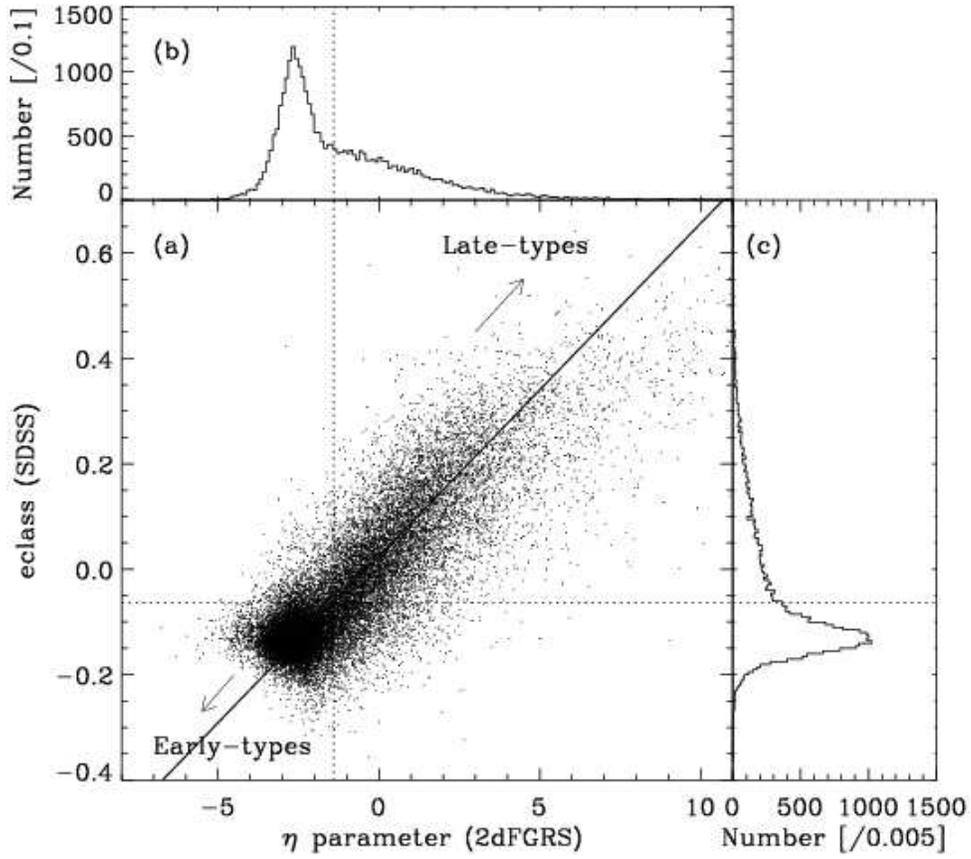}
\centering
\caption{(a) Comparison of \texttt{eclass} parameter in SDSS with $\eta$
parameter in 2dFGRS,
(b) histogram for $\eta$ parameter, and
(c) histogram for \texttt{eclass} parameter.
Solid line indicates the best linear fit for \texttt{eclass} ($<0.09$) and $\eta$ ($<1.1$).
The dotted vertical line denotes the division value $\eta=-1.4$ in 2dFGRS
for galaxy subsamples, and dotted horizontal line indicates
\texttt{eclass}=$-0.0640$ in SDSS, which is equivalent to the division value $\eta=-1.4$ in 2dFGRS.
}\label{fig-spcl}
\end{figure}

To investigate the difference of galaxy orbits among subsamples,
we divide the galaxies into early-type and late-type galaxies
using spectroscopic parameters provided by each survey.
For 2dFGRS galaxies, $\eta$ parameter, which is a linear combination
of emission and absorption components within the spectrum
derived from the Principal Component Analysis,
is assigned to each galaxy \citep{mad02}.
This parameter denotes the ratio of the
present to the past star-formation activity in each galaxy.
\citet{mad02} showed that the $\eta$ parameter correlates well with galaxy morphology
using the Kennicutt Atlas \citep{ken92} (see their Fig. 4).
They reported that the galaxies with $\eta<-1.4$ are corresponding to E/S0 and Sa
and those with $\eta>-1.4$ to Sb and Scd.
Similarly, for SDSS galaxies, \texttt{eclass} parameter, which is
a projection of the first three principal components of the spectrum,
is assigned to each galaxy.
This parameter has values from about $-0.35$ (corresponding to early-type galaxies) 
  to 0.5 (late-types), which was used to classify SDSS galaxies (e.g., \citealt{ber03,ber06}).
In Figure \ref{fig-spcl}, 
  we plot the \texttt{eclass} parameter in SDSS versus the $\eta$ parameter in 2dFGRS
  in order to determine the \texttt{eclass} value corresponding to the division value of $\eta=-1.4$.
It is seen that two parameters correlate well and the majority of galaxies are
  located around ($\eta$, \texttt{eclass})=($-2.65$,$-0.135$).
We determined the relation equation between two parameters
  using the ordinary least-square bisector method \citep{iso90}. 
The fit was done for the galaxies with $\eta<1.1$ and \texttt{eclass}$<0.09$.
\begin{equation}
{\rm \texttt{eclass}}_{\rm SDSS} = 0.0631 (\pm0.0004)~{\eta}_{\rm 2dFGRS}+0.0244 (\pm0.0008) \label{eq-trans}
\end{equation}

This relation indicates that the division value of $\eta=-1.4$ in 2dFGRS is equivalent
to that of \texttt{eclass}=$-0.0640$ in SDSS. 
Therefore, we classified the galaxies into early-type galaxies if
  (i) $\eta<-1.4$ when the galaxy is surveyed only in 2dFGRS,
  (ii) \texttt{eclass}$<-0.0640$ when the galaxy is surveyed only in SDSS, and
  (iii) \texttt{eclass}$<-0.0640$ and $\eta<-1.4$ when the galaxy is surveyed both in SDSS and 2dFGRS.
The rest were classified into late-type galaxies.
Some galaxies without estimation of $\eta$ parameter ($\eta=-99.9$)
were not used for the analysis of subsamples.

\subsection {Cluster Sample and Galaxy Membership in Clusters}\label{mem}
We used the Abell catalog of galaxy clusters \citep{aco89} to
identify cluster galaxies in the survey data. Among the Abell
clusters, we selected those that have known spectroscopic
redshifts in the NASA/IPAC Extragalactic Database (NED). Finally,
we selected 731 and 230 clusters located within the survey regions
of SDSS and 2dFGRS, respectively, as a sample of clusters for
further analysis.

In order to determine the membership of galaxies in a cluster, we used
the ``shifting gapper'' method of \citet{fadda96} as used also for
the study of global rotation of galaxy clusters \citep{hl07}.
In the plot of radial velocity versus clustercentric distance of galaxies for a
given cluster, we selected the member galaxies using a
velocity gap of 950 km s$^{-1}$ and a distance bin of 0.2 Mpc
shifting along the distance from the cluster center.
We used a larger bin width if the number of galaxies in a
bin was less than 15. We applied this method to the galaxies within
the radius at which the distance between adjacent galaxies is
larger than 0.1 Mpc. If there are no adjacent galaxies
with $>0.1$ Mpc, we stopped the procedure at the radius of 3.5 Mpc.
We iterated the procedure until the number of
cluster members is converged. Finally, we selected 113 galaxy
clusters in which the number of member galaxies is greater than or
equal to 40 for further analysis.

Since it is necessary to determine the galaxy orbits
using cluster mass profiles from X-ray data, we selected  21 galaxy clusters
whose X-ray mass profiles are available in the literature (e.g., \citealt{rb02,san03,dem03,bm06}).
Then, we rejected 9 clusters (A1775, 2052, 2063, 2142, 2147, 2244, 2255, 4038, and 4059)
that appear to be in the stage of interacting or merging
in the plot of galaxy velocity versus clustercentric distance,
and 2 clusters (A119 and A1656) whose survey coverages are not complete.
Although A85 was partially surveyed, we included it into our sample
since the uncovered region is only the small outer region of the entire cluster.
Finally, we obtained a sample of 10 clusters
that will be used for investigating the galaxy orbits.

\begin{figure}
\includegraphics [width=135mm]{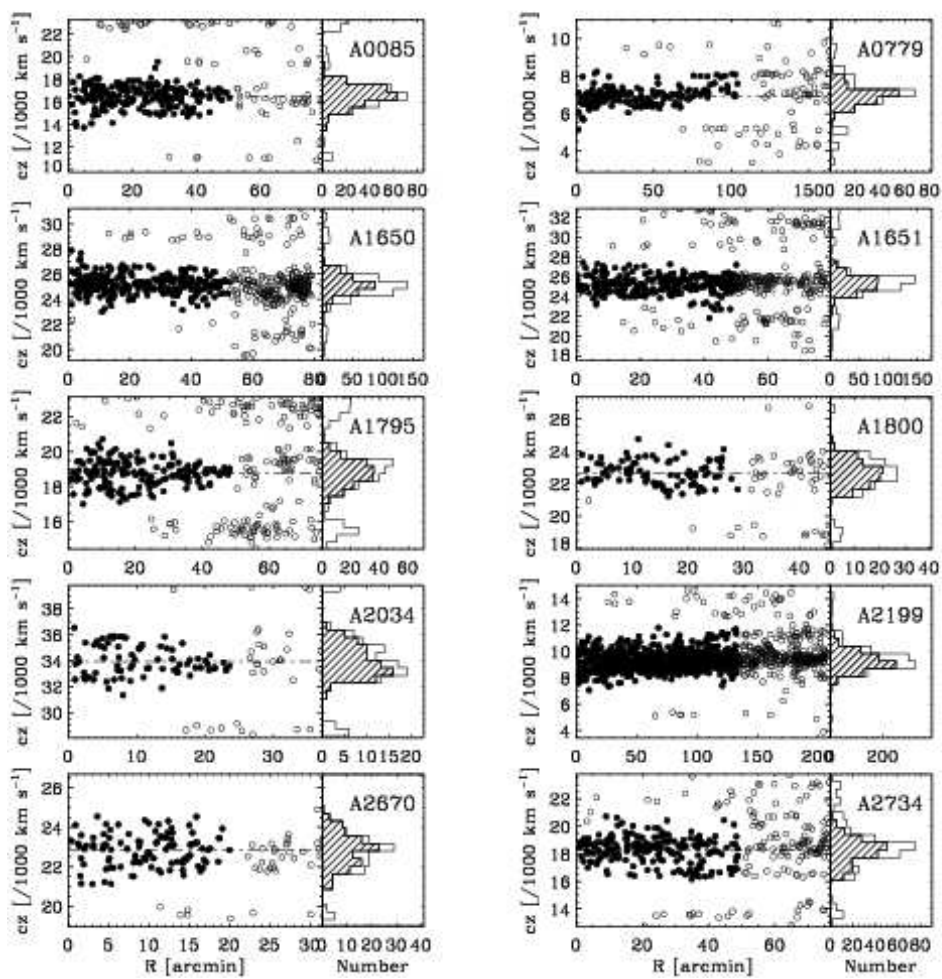}
\centering
\caption{Radial velocity vs. clustercentric distance of galaxies
and the velocity distribution for our sample clusters. Filled circles
indicate the galaxies selected as cluster members, while open
circles the galaxies not selected as cluster members.\ The
horizontal dot-dashed lines indicate the systemic velocity of the
clusters determined in Table \ref{tab-cand}. The velocity
distributions for the member galaxies are shown by hatched
histograms, and those for all of the observed galaxies by open
histograms.} \label{fig-member}
\end{figure}

\begin{figure}
\centering
\includegraphics [width=135mm]{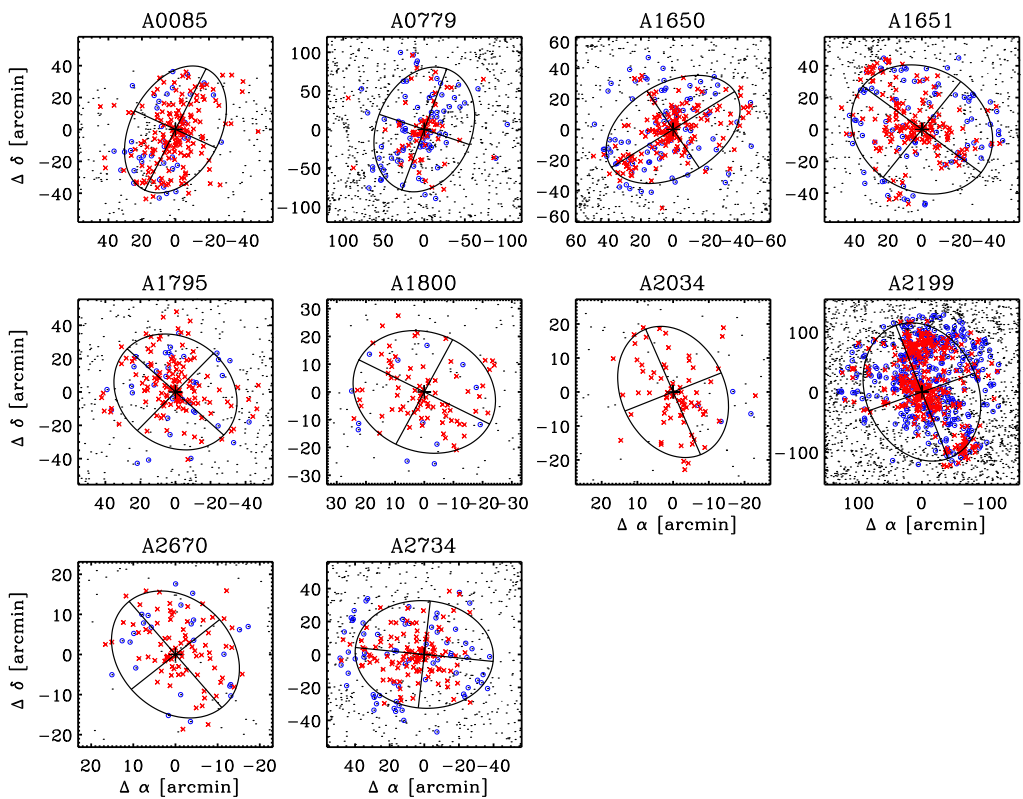}
\caption{Spatial distribution of cluster galaxies with measured
velocities for our sample clusters. Early-type galaxies
are plotted with crosses, while late-type galaxies are
with open circles. All the observed
galaxies are represented by dots. The ellipses indicate the
(twice enlarged) dispersion ellipse, and solid lines denote the
major and minor axis of the dispersion ellipse.
}\label{fig-spvel}
\end{figure}

Table \ref{tab-cand} lists our sample clusters with Abell identification, right ascension and declination,
Bautz-Morgan (B-M) type, the survey in which the cluster is covered,
the redshift derived in this study (the biweight location of \citealt{beers90}),
the physical extent corresponding to one arcmin,
and the velocity dispersion (the biweight scale of \citealt{beers90}) and
the number of galaxies for all, early-type, and late-type galaxies.
Our sample clusters are found from $z=0.023$ to $z=0.113$ and have 78$-$754 member galaxies.
The number of early-types is usually larger than that of late-types in clusters,
and eight clusters have B-M type of I or II.
Figure \ref{fig-member} shows plots of radial velocity as a function of clustercentric distance of galaxies
and the velocity distributions for the 10 clusters.
In Figure \ref{fig-spvel}, we show the spatial distribution of cluster galaxies
  with measured velocities. 
For most clusters, early-types are centrally concentrated, while late-types are not.

\begin{figure}
\centering
\includegraphics [width=135mm]{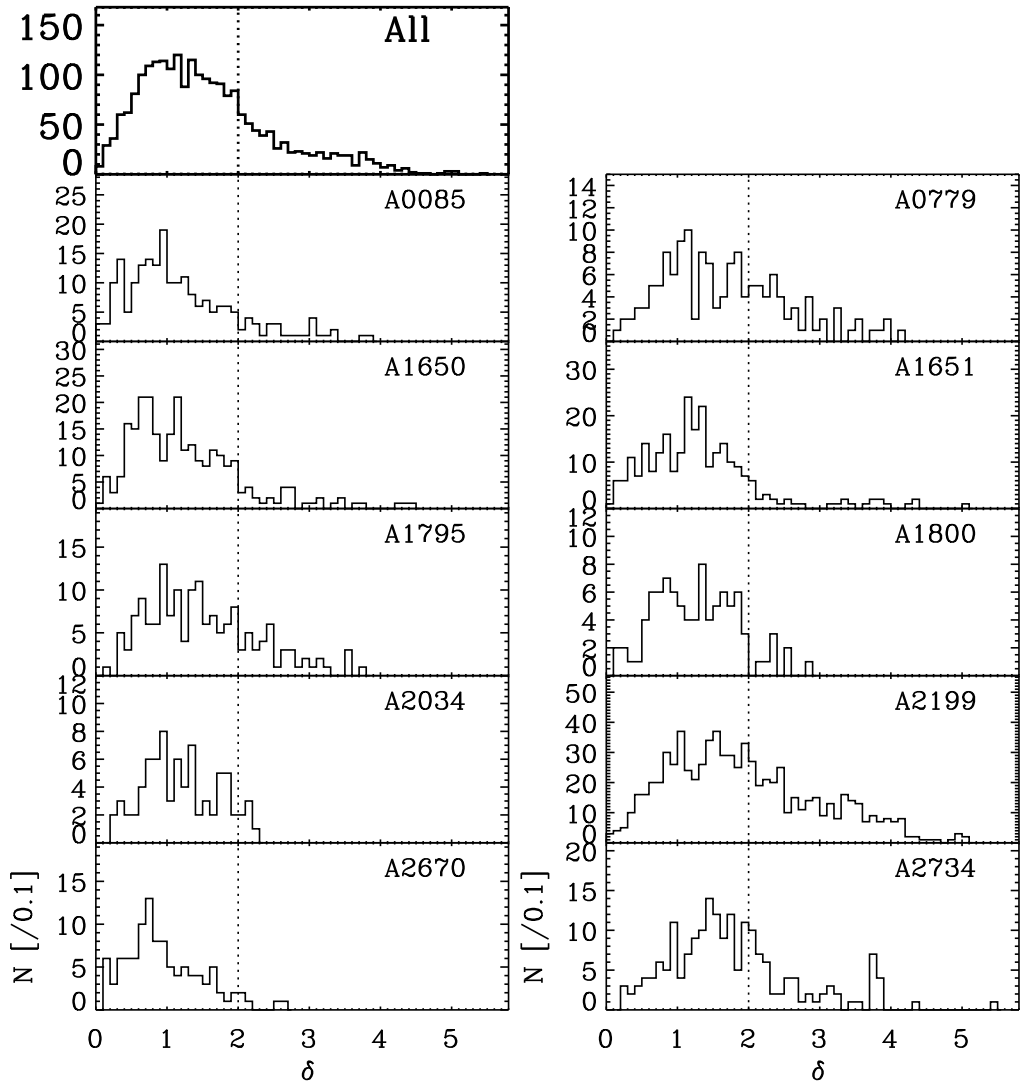}
\caption{Distribution of the substructure parameter $\delta$ for our sample clusters.
The dotted vertical lines indicate the selection criteria for the galaxies
  in the cluster main body.
}
\label{fig-delta}
\end{figure}

Since the presence of substructure in clusters can affect
  the determination of the VDPs and the galaxy number density profiles,
  it is necessary to secure the galaxy sample outside the substructure 
  in order to determine the galaxy orbits properly.
Therefore, we show, for our sample cluster, the distribution of $\delta$ that indicates
  local deviations from the systemic velocity ($v_{\rm sys}$) and dispersion ($\sigma_p$) of 
  the entire cluster in Figure \ref{fig-delta} (see \S \ref{sub} for the explanation of $\delta$ in detail).
We selected the galaxies with $\delta\leq2.0$ that 
  are regarded as the galaxies in the cluster main body for the orbit analysis,
  and compared the results for the different choice of $\delta$ value in \S \ref{galsample}.

\section{Observed Properties of Galaxy Clusters}\label{property}
\subsection {X-ray Mass Profile}\label{xmass}

\begin{figure}
\centering
\includegraphics [width=135mm]{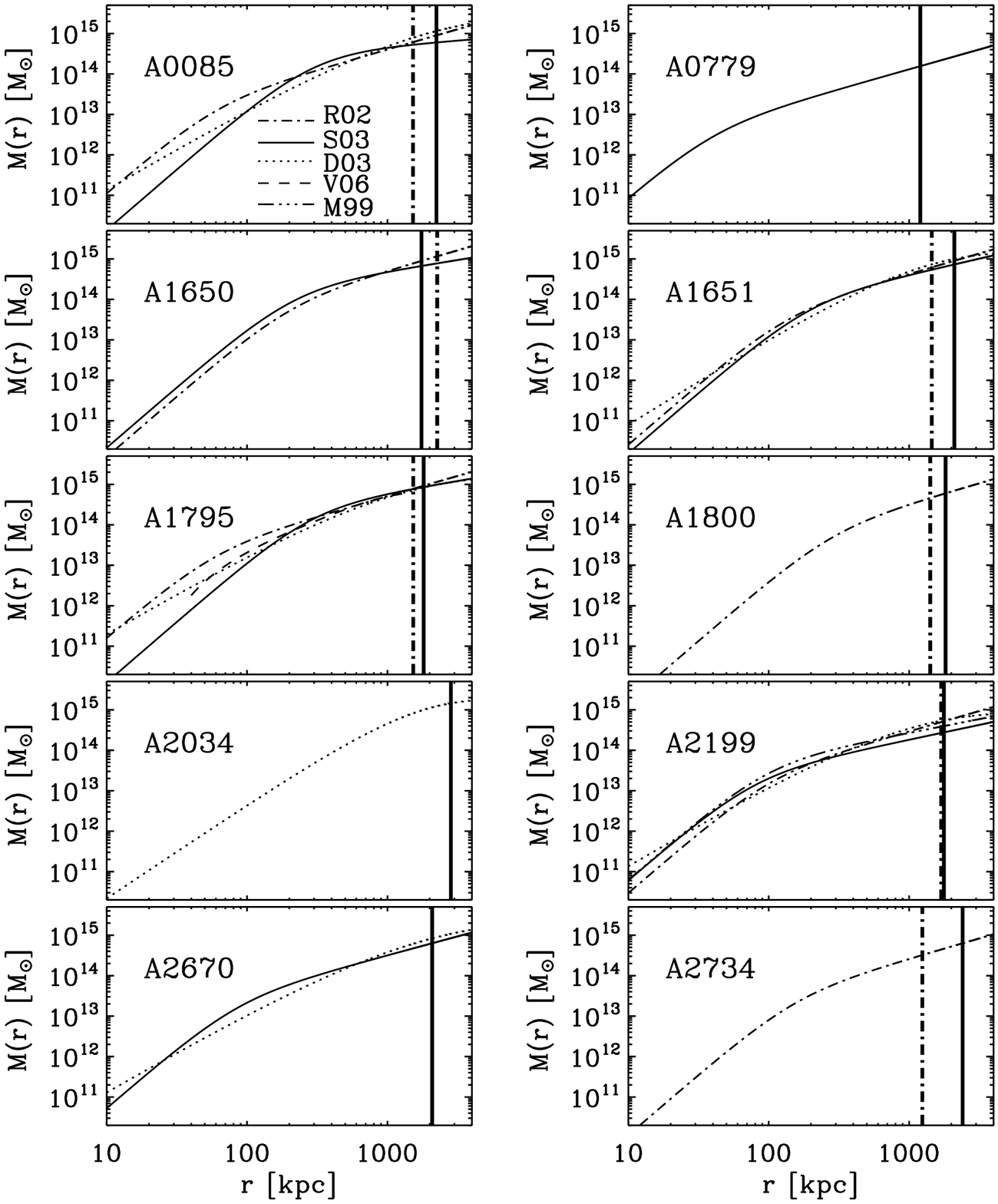}
\caption{Total mass profiles for our sample clusters.
Dot-dashed lines from \citet[R02]{rb02},
solid lines are from \citet[S03]{san03}, 
dotted lines from \citet[D03]{dem03},
dashed lines from \citet[V06]{vik06}, and
dot-dot-dashed lines from \citet[M99]{mar99}.
The vertical dot-dashed and solid lines indicate the outer significance radius, $r_X$,
  in \citet{rb02} and $r_{200}$ radius computed in this study, respectively.
}
\label{fig-mass}
\end{figure}

We present X-ray mass profiles for our sample clusters in Figure \ref{fig-mass}.
It appears that mass profiles from various references agree well as a whole in each cluster.
In Table \ref{tab-xray}, we list the X-ray luminosity and the parameters for
  the X-ray mass profile.
The first and second columns give Abell identification and the X-ray luminosity, respectively.
The third and fourth columns are $\beta$ parameter and core radius ($r_c$)
  derived from the standard $\beta$-model of cluster gas density profile, respectively (e.g., \citealt{cf76}). 
The fifth and sixth columns are isothermal gas temperature ($T_X$) in \citet{rb02}, and
  central gas temperature [$T(0)$] in \citet{san03}, respectively.
The seventh column gives the polytropic index ($\gamma$) used in equation (\ref{eq-temp}), and
  final column gives the reference of these parameters adopted in this study.
Then, the gravitational cluster mass $M_{\mathrm {tot}}(r)$ is determined
  under the assumption of hydrostatic equilibrium through the equation,
\begin{equation}
M_{\mathrm {tot}}(r) = - {{k T(r) r} \over {G \mu m_p}} \left(
{{{\mathrm {d~ln} \rho (r)} \over {\mathrm {d~ln} r}} + {{\mathrm {d~ln} T(r)} \over {\mathrm {d~ln} r}}} \right),
\label{eq-mass}
\end{equation}

where $k$ is the Boltzmann constant, 
  $G$ is the gravitational constant, 
  $\mu$ ($=0.61$) is the mean molecular weight, 
  $m_p$ is the proton mass.
The gas density profile is given by,
\begin{equation}
\rho(r)=\rho(0)\left[1+\left(\frac{r}{r_{c}}\right)^{2}\right]^
 {-\frac{3}{2}\beta}.
\label{eq-gasden}
\end{equation}

For the clusters in \citet{rb02},
  we use isothermal gas temperature ($T_X$) as $T(r)$.
For the clusters in \citet{san03}, we use non-isothermal gas temperature profile
  that is linked to the gas density via a polytropic equation of state,
\begin{equation}
T(r)=T(0)\left[1+\left(\frac{r}{r_{c}}\right)^{2}\right]^
 {-\frac{3}{2}\beta \left(\gamma - 1\right)},
\label{eq-temp}
\end{equation}
where $\gamma$ is the polytropic index and $r_c$ and $\beta$ are as defined previously.
For A2034, the X-ray mass profile is available in \citet{dem03}. 
However, they present a different form of gas density profile from the standard $\beta$-model, 
  so those parameters were not included in Table \ref{tab-xray}.

It is worth noting that the X-ray mass profiles may not be well determined
  outside the outer significance radius of the cluster, $r_X$,
  because the X-ray source count drops below the Poissonian $1\sigma$ error at $r>r_X$ \citep{rb02}.
It is also noted that the Jeans equation may not be applicable 
  beyond a radius of $r_{200}$ (usually called virial radius),
  because the dynamical equilibrium of the cluster is not guaranteed at this region.
Therefore, careful interpretation is needed beyond $r_X$ and $r_{200}$.
We draw the vertical lines at the radius of $r_X$ and $r_{200}$ in Figures \ref{fig-mass},
  \ref{fig-orbit1}, \ref{fig-orbit2}, \ref{fig-orbit3}, \ref{fig-orbitab}, and \ref{fig-orbitem}.
The radius $r_{200}$ that contains an overdensity 200$\rho_{\rm c}$
  where $\rho_{\rm c}$ is the critical density of the Universe,
  is computed for each cluster using the equation in \citet{car97a}:
\begin{equation}
r_{200}= \frac{3^{1/2}\sigma_{\rm cl}}{10 H(z)},
\end{equation}
where $\sigma_{\rm cl}$ is a velocity dispersion of the cluster and
  the Hubble parameter at $z$ is
  $H^2(z)=H^2_0 [\Omega_M(1+z)^3 +\Omega_k(1+z)^2+\Omega_\Lambda]$ \citep{pee93}.
$\Omega_M$, $\Omega_k$, and $\Omega_\Lambda$ are dimensionless density parameters. 

\subsection {Galaxy Number Density Profile}

We derived the galaxy number density profile for each cluster
  using the member galaxies with $\delta\leq2.0$ selected in \S \ref{mem}. 
Since SDSS and 2dFGRS are magnitude limit surveys ($r_{\rm pet}<17.77$ for SDSS 
 and ${\rm b_J}<19.45$ for 2dFGRS) and the spectroscopic sample of galaxies are 
 nearly complete within the magnitude limit,
 we can derive the galaxy number density profile for each cluster 
 using spectroscopically selected member galaxies.

\begin{figure}
\centering
\includegraphics [width=125mm]{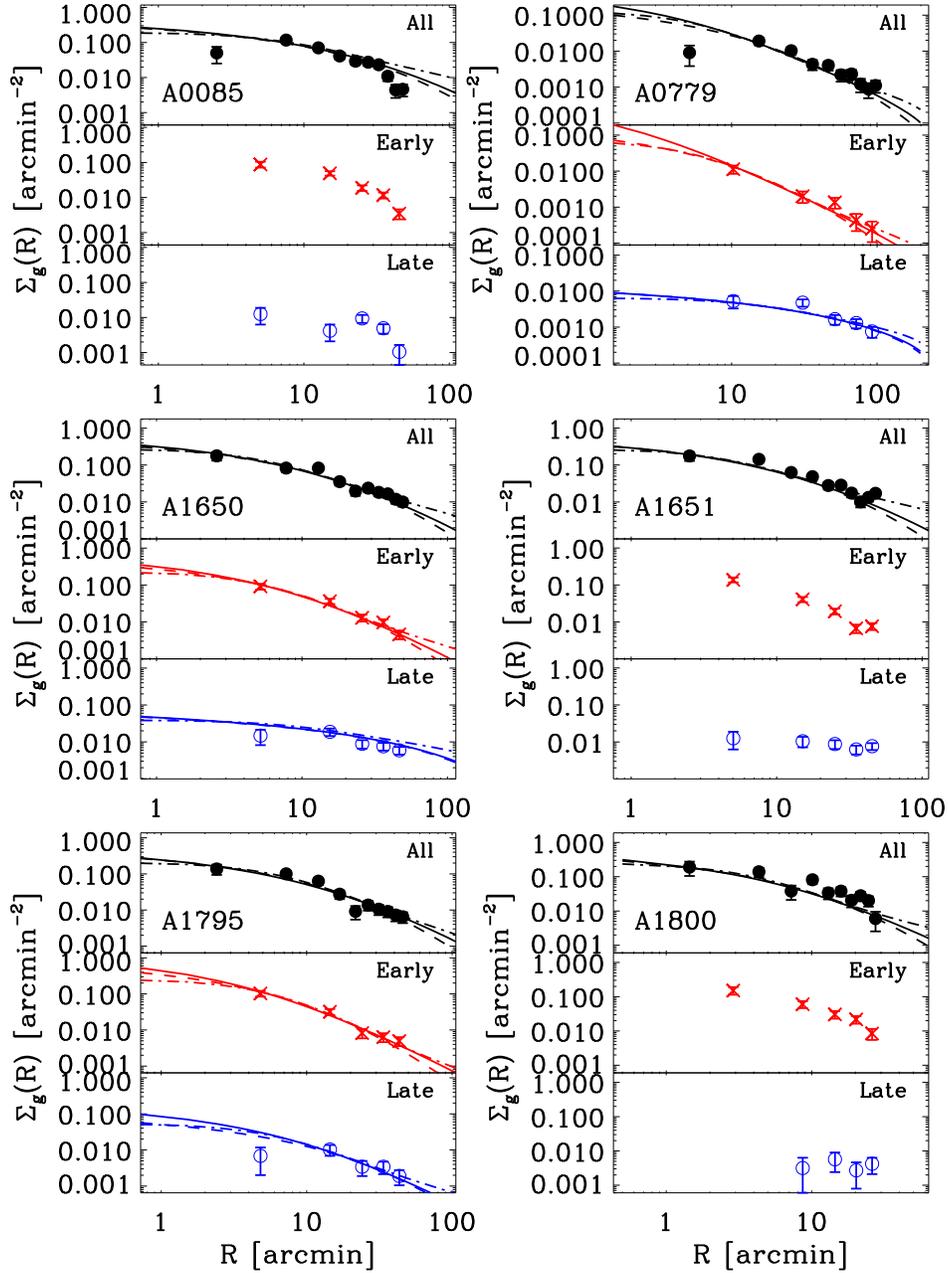}
\caption{Projected number density profiles of all (top panel),
early-type (middle panel), and late-type (bottom panel) galaxies with $\delta\leq2.0$
for A85, 779, 1650, 1651, 1795, and 1800. The solid, dashed, and dot-dashed
lines indicate the projection of the best fits using NFW,
Hernquist, and KEK05 profiles, respectively, for each sample.
}\label{fig-numden1}
\end{figure}

\begin{figure}
\centering
\includegraphics [width=135mm]{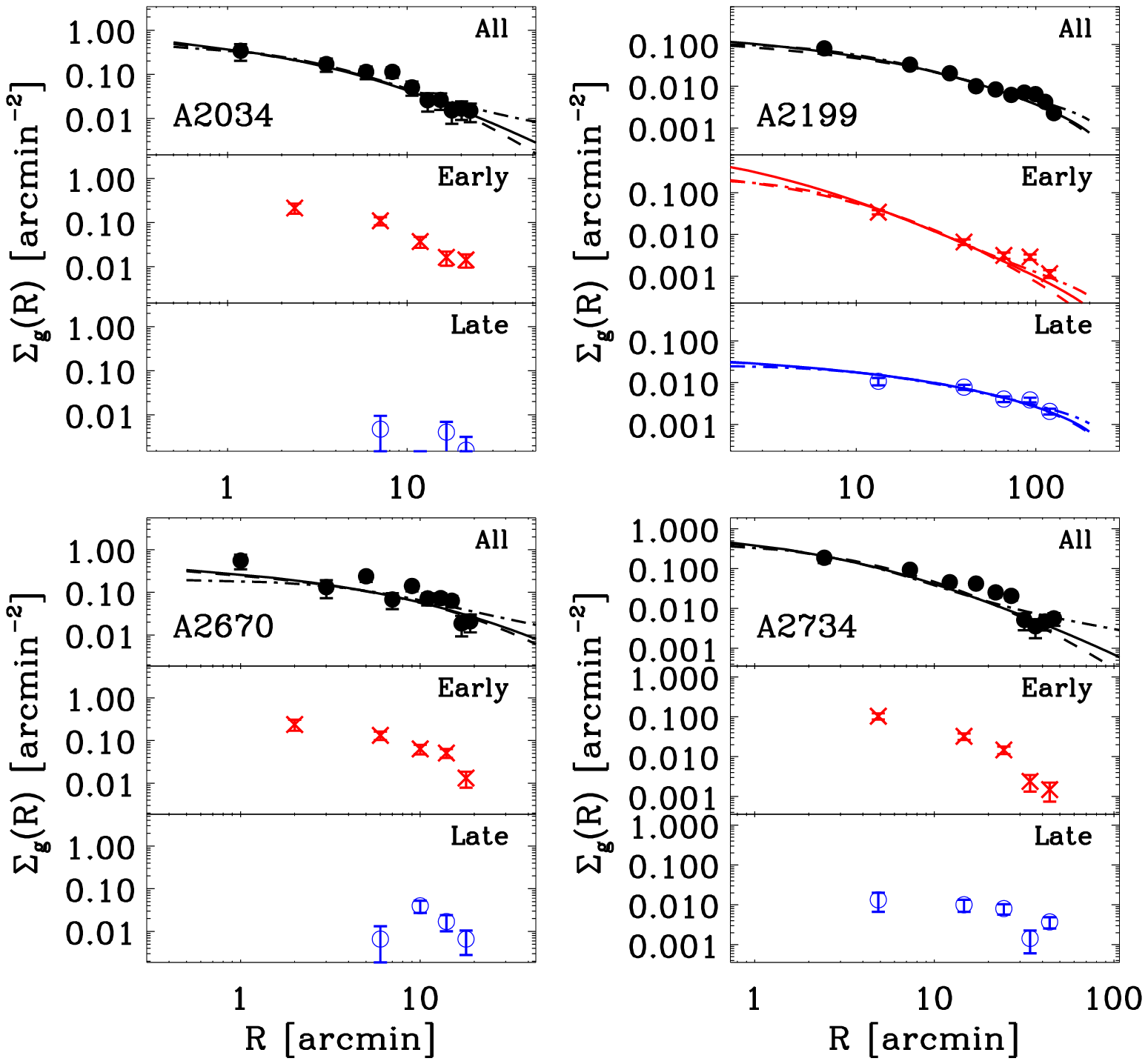}
\caption{Same as Fig. \ref{fig-numden1}, but for A2034, 2199,
  2670, and 2734. }\label{fig-numden2}
\end{figure}

We display the galaxy number density profiles of all, early-type, and
  late-type galaxies for each cluster in Figures \ref{fig-numden1} and
  \ref{fig-numden2}. 
We de-projected the observed number
  density using the method in \citet{mcl99}, and fitted the de-projected
  density profile with those of \citet[NFW]{nfw97}
  and \citet{her90}, which are represented by,
\begin{equation}\label{eq-den}
\nu_{\rm g} (r)=\nu_0(r/r_s)^{-1}(1+r/r_s)^{-\alpha},
\end{equation}

where $\nu_{\rm g}(r)$ is a three dimensional density profile of the cluster galaxies,
  $r_s$ is a scale radius, and $\alpha=2$ for the NFW profile
  and $\alpha=3$ for the Hernquist profile.
We also fitted the de-projected density profile with that of
\citet[KEK05]{kek05}, which is derived from the isotropic and isothermal galaxy
orbits in clusters under the NFW distribution of dark matter.
It is represented by,
\begin{equation}\label{eq-den2}
\nu_{\rm g} (r)=\nu_0 \left[ \frac{(1+r/r_s)^{(r_s/r)}}{e}\right]^{\eta_g}
\end{equation}
where the dimensionless parameter $\eta_g$ is defined by
\begin{equation}\label{eq-eta}
\eta_g \equiv \frac{2G M_0}{r_s\sigma_r^2}.
\end{equation}

We used as a characteristic mass $M_0$,
  X-ray mass estimate shown in Figure \ref{fig-mass} at a radius of $r_{200}$,
  and used as a one-dimensional velocity dispersion $\sigma_r$,
  observed velocity dispersion $\sigma_p$ shown in Table \ref{tab-cand}.

The solid, dashed, and dot-dashed
lines represent the projected best fit curves of the NFW,
Hernquist, and KEK05 profiles, respectively. We did not fit the profiles that
give unstable de-projection, and did not include them in Figures \ref{fig-numden1} and \ref{fig-numden2}.
For the clusters (A1650, 1651, and 2670) surveyed in both SDSS and 2dFGRS, we used
one data set (2dFGRS for A1650 and A1651, SDSS for A2670) for deriving number density profile
since the spatial coverage of SDSS or 2dFGRS is not complete depending on clusters.
For A85, the fit was done using only the galaxies in the
  inner region ($R<26\arcmin$) because of incomplete coverage
  (see Fig. \ref{fig-spvel}).

The fitting results for our sample are summarized in Table \ref{tab-numden}.
It is found that the scale radius $r_s$ of late-types
  in all profiles 
  is larger than that of the early-types, 
  showing that the latter is more concentrated toward
  the cluster center than the former. 
The steeper density profile of early-types compared with late-types is a clear indication of the
  morphology-density relation (e.g., \citealt{dre80}).

\begin{figure}
\centering
\includegraphics [width=135mm]{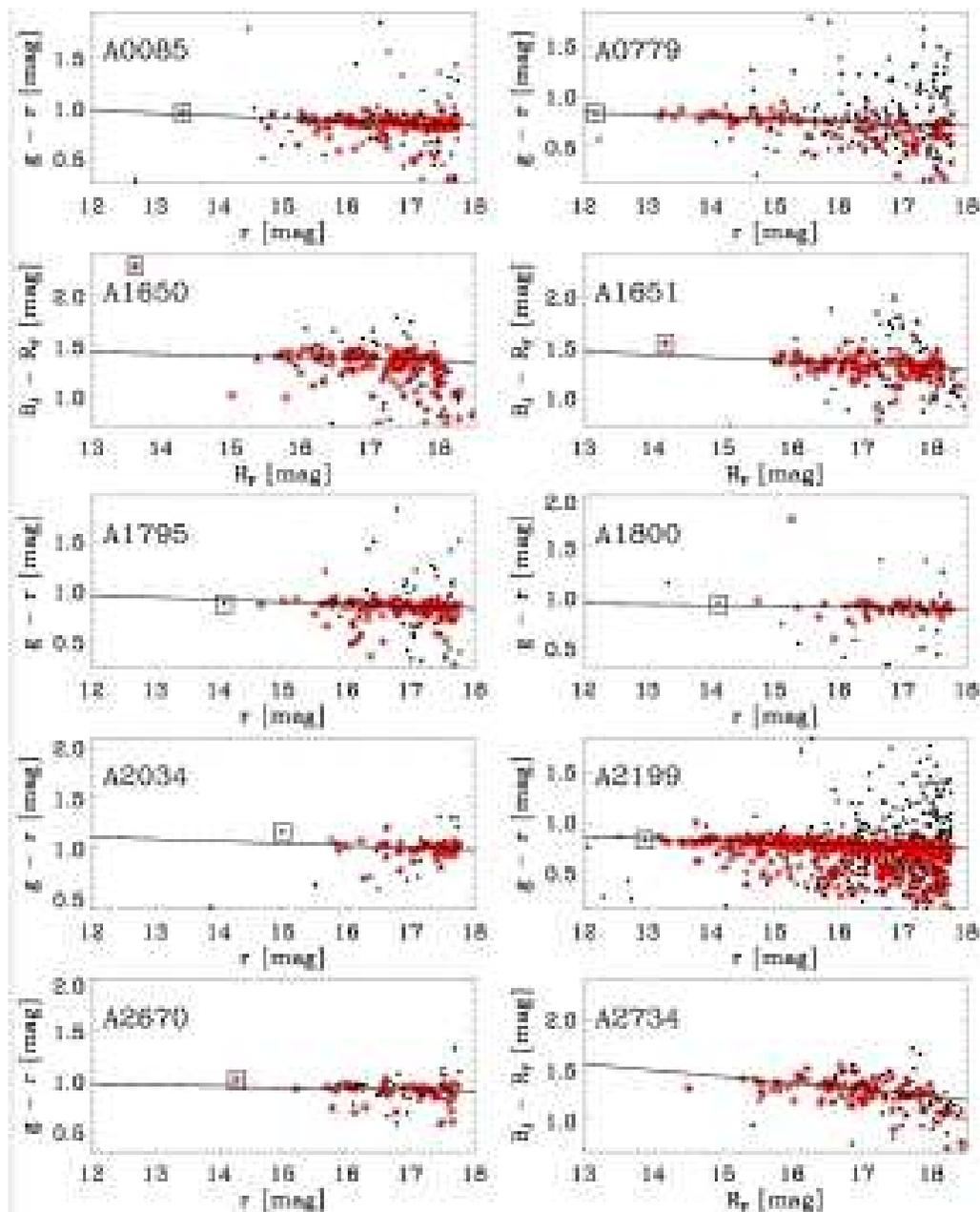}
\caption{Color-magnitude diagrams for our sample clusters.
Filled circles indicate the photometric sample of galaxies
within the $R_{max}$/2 from the cluster center,
while open circles the spectroscopically selected member galaxies in the same region.
Solid lines indicate the best linear fit of the CMR for early-type galaxies.
BCGs are marked by open squares.
}\label{fig-cmr}
\end{figure}

To test the effect of the incompleteness of spectroscopic sample
  on the galaxy number density profiles,
  we investigate the color magnitude diagram of the galaxies 
  in the photometric and spectroscopic samples.
It is well known that optical colors of early-type galaxies are 
  strongly correlate with absolute magnitudes (color-magnitude relation, CMR), 
  in the sense that the brighter galaxies are likely to be redder
  than the fainter galaxies \citep{baum59,faber73,vs77}.
In Figure \ref{fig-cmr},
  we plot the $g-r$ versus $r$ for the SDSS cluster galaxies and the $B_J-R_F$ versus $R_F$
  for the 2dFGRS cluster galaxies.
The cluster galaxies within the $R_{max}$/2 from the cluster center
  selected in \S \ref{mem} are shown in company with
  the photometric sample of galaxies without measured velocities in the same region.
$R_{max}$ is the largest clustercentric distance of the member galaxies, 
  and varies depending on clusters.
It shows that the colors correlate well with the observed magnitudes.
The linear fits derived from repeated one sigma clipping using early-type galaxies
selected in \S \ref{gal} are overlaid.
The average of the CMR slopes for our sample clusters is found to be
$-0.019\pm0.006$ for SDSS clusters and $-0.037\pm0.024$ for
2dFGRS clusters,
which is consistent with the results of previous studies
based on the same data \citep{prop04,hogg04,gal06,agu07}.
It appears that most bright ($r<14$ mag) galaxies following the CMR
in the photometric sample are selected as member galaxies,
implying that the incompleteness of spectroscopic sample in the cluster center
does not affect the determination of
the projected galaxy number density in Figures \ref{fig-numden1} and \ref{fig-numden2}.

\subsection {Velocity Dispersion Profile}\label{vdp}
\begin{figure}
\centering
\includegraphics [width=115mm]{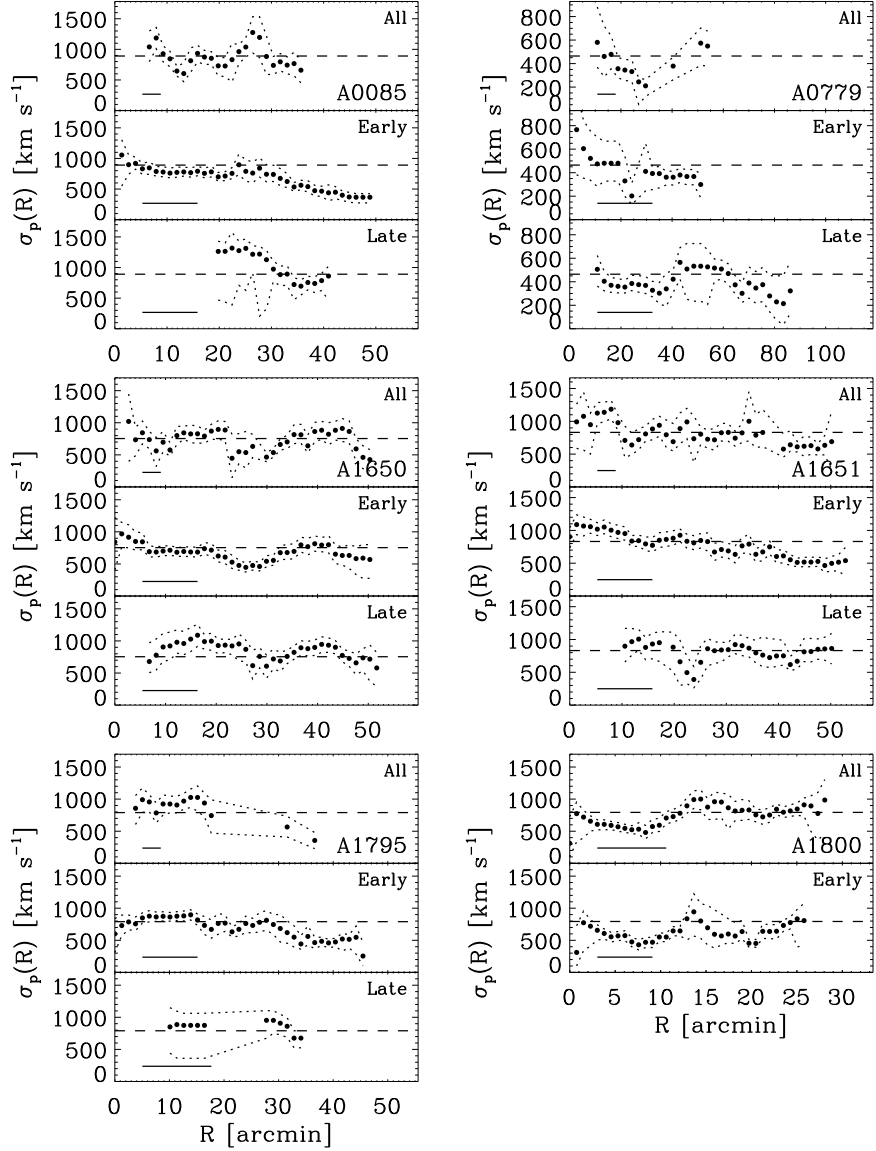}
\caption{Velocity dispersion profiles of all (top panel),
early-type (middle panel), and late-type (bottom panel) galaxies with $\delta\leq2.0$
for A85, 779, 1650, 1651, 1795, and 1800. Filled circles indicate
the velocity dispersion ($\sigma_p$) at each point. The dispersion
is calculated using the galaxies within the moving radial bin that
is represented by a horizontal errorbar in each panel. The dotted
lines denote 68\% confidence interval on the calculation of
velocity dispersion. The dashed horizontal line indicates the
global value of velocity dispersion using all galaxies in each
cluster.}\label{fig-disp1}
\end{figure}

\begin{figure}
\includegraphics [width=135mm]{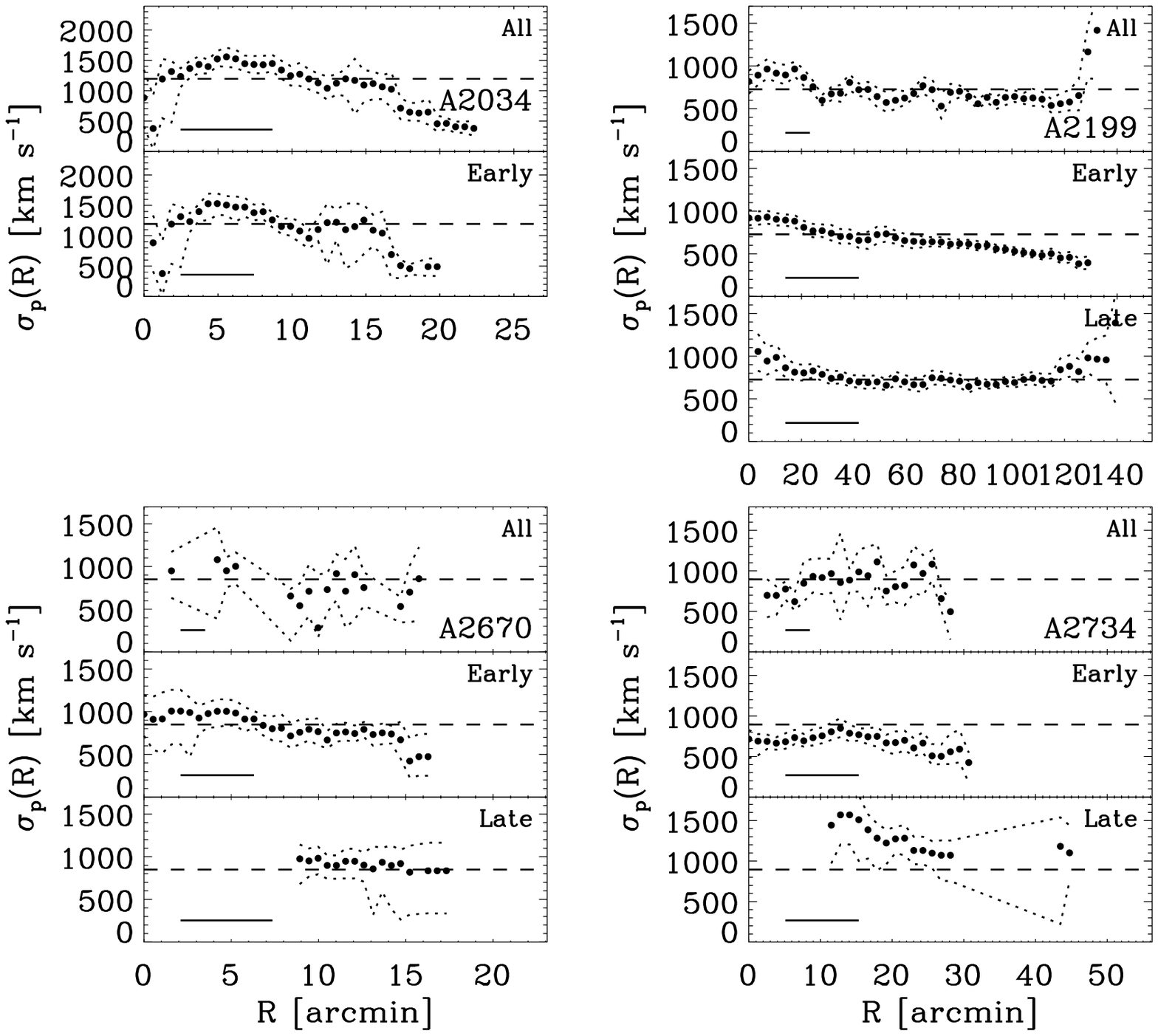}
\caption{Same as Fig \ref{fig-disp1}, but for A2034, 2199,
2670, and 2734. }\label{fig-disp2}
\end{figure}

\begin{figure}
\centering
\includegraphics [width=95mm]{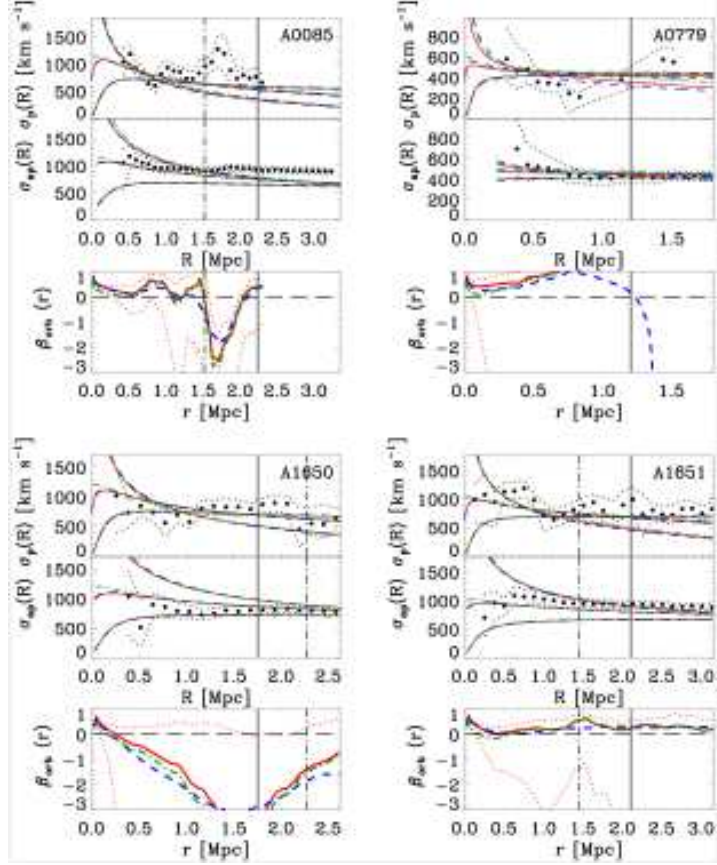}
\caption{Velocity dispersion profile (top panel) and the aperture
VDP (middle panel), velocity anisotropy profile (bottom panel) for
all galaxies of A85, 779, 1650, and 1651. Filled circles in top
panels represent the measured VDP shown in Fig. \ref{fig-disp1},
and those in middle panel denote the measured aperture VDP.
Associated, dotted lines represent 68\% confidence interval on the
calculation of velocity dispersion. Three smoothly curved lines in
top and middle panels, from a radial anisotropy to the tangential
one (from top to bottom, $\beta_{\rm orb}$= 0.99, 0, and $-$99),
represent the calculated dispersion profiles using galaxy number
density profile of NFW (solid lines), Hernquist (dashed lines), and KEK05 (dot-dashed lines).
Velocity anisotropy profiles (VAPs) determined in \S \ref{method2} in the
  bottom panel are represented by solid (NFW), dashed (Herquinst),
  and dot-dashed line (KEK05).
The errors of VAPs for NFW are shown by dotted lines. 
The vertical dot-dashed and solid lines indicate the outer significance radius, $r_X$,
  in \citet{rb02} and $r_{200}$ radius computed in this study, respectively.
}\label{fig-orbit1}
\end{figure}

\begin{figure}
\centering
\includegraphics [width=135mm]{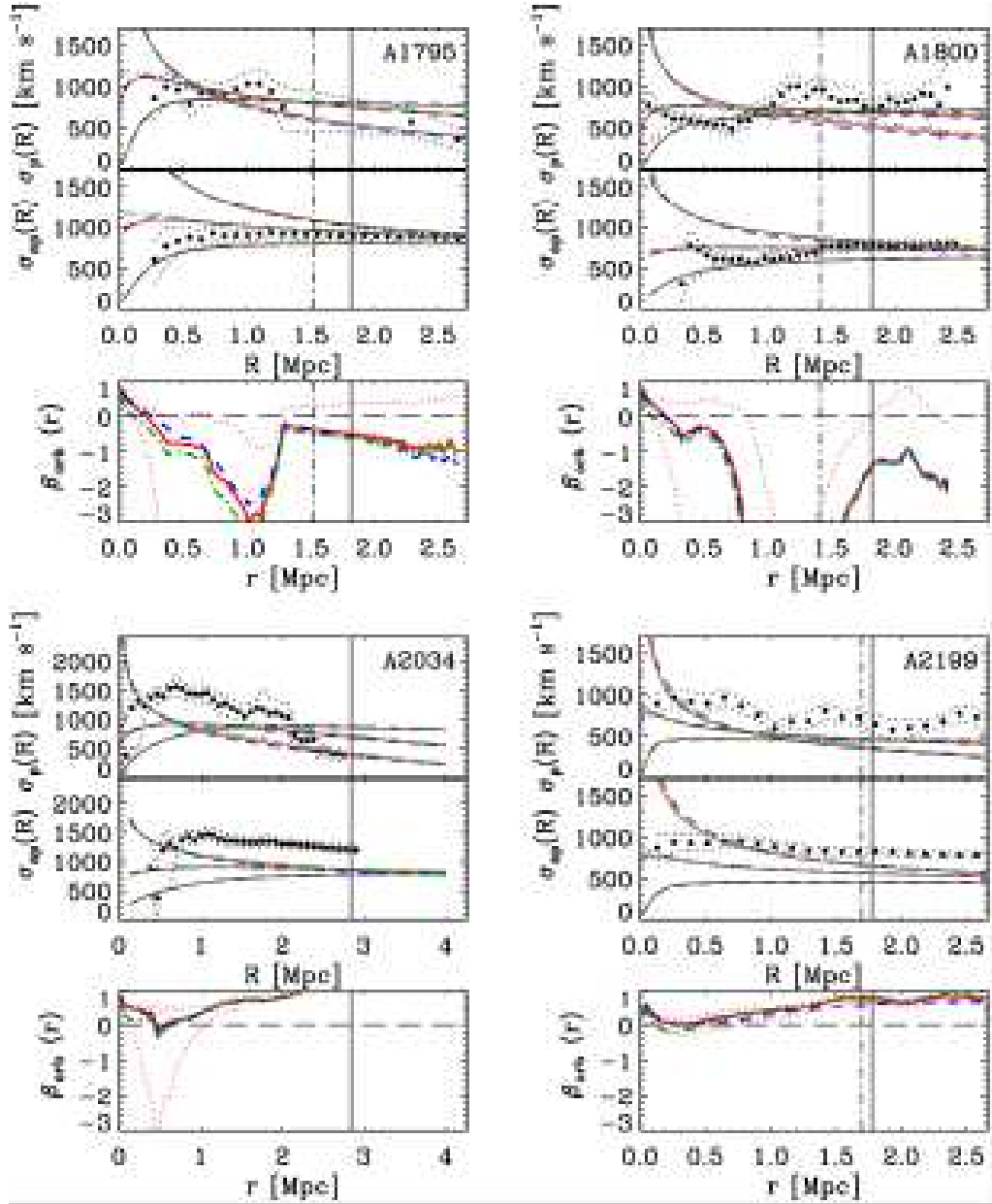}
\caption{Same as Fig. \ref{fig-orbit1}, but for A1795, 1800, 2034, and 2199.
}\label{fig-orbit2}
\end{figure}

\begin{figure}
\centering
\includegraphics [width=135mm]{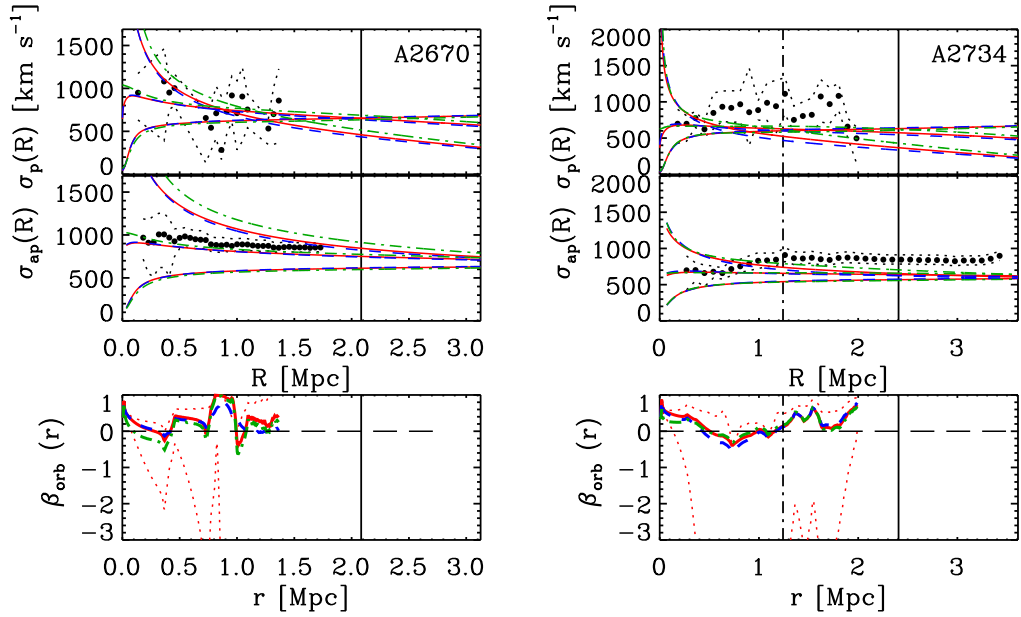}
\caption{Same as Fig. \ref{fig-orbit1}, but for A2670 and A2734. }\label{fig-orbit3}
\end{figure}

\begin{figure}
\centering
\includegraphics [width=135mm]{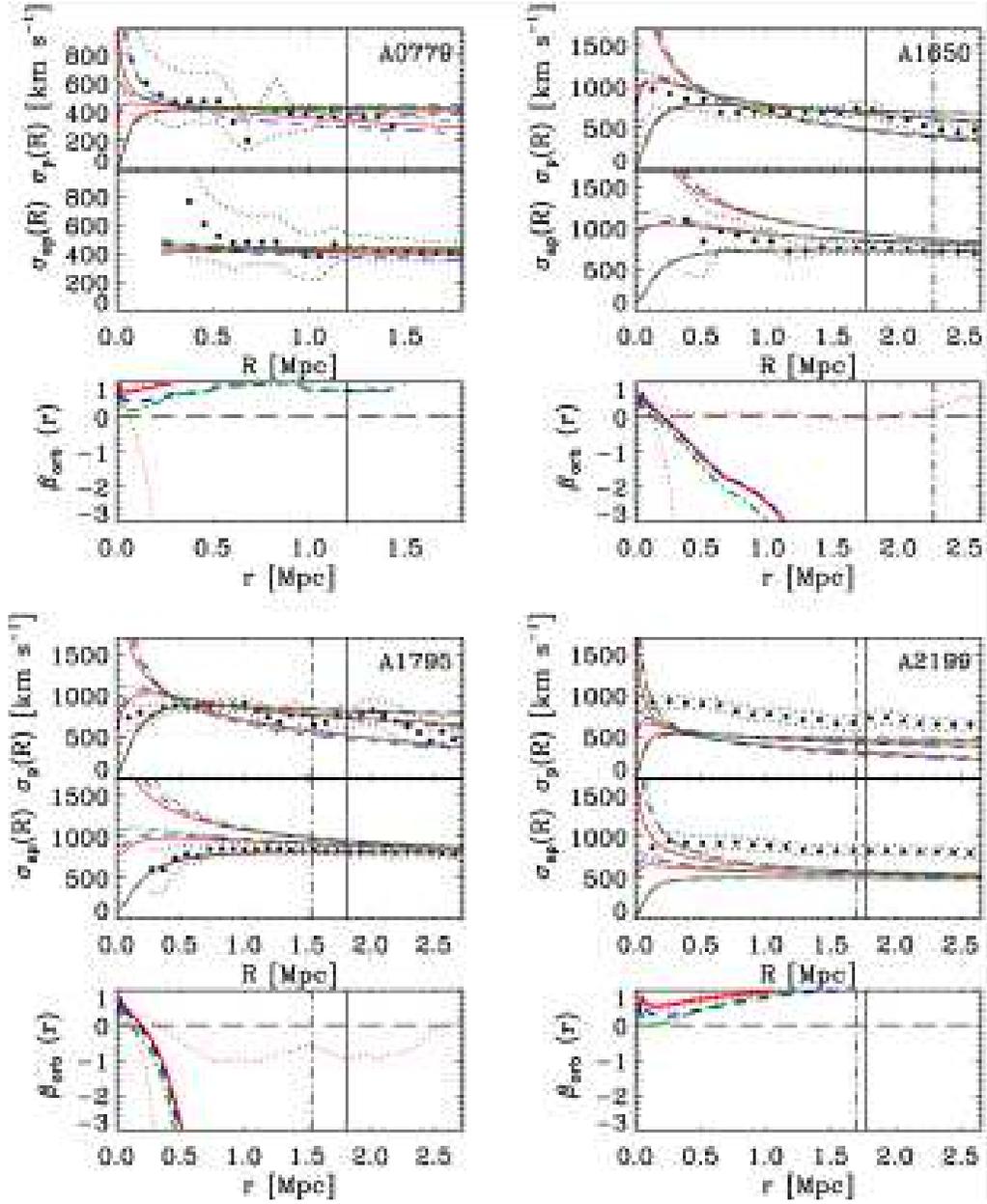}
\caption{Same as Fig. \ref{fig-orbit1}, but for early-type galaxies in
A779, 1650, 1795, and 2199. }\label{fig-orbitab}
\end{figure}

\begin{figure}
\centering
\includegraphics [width=135mm]{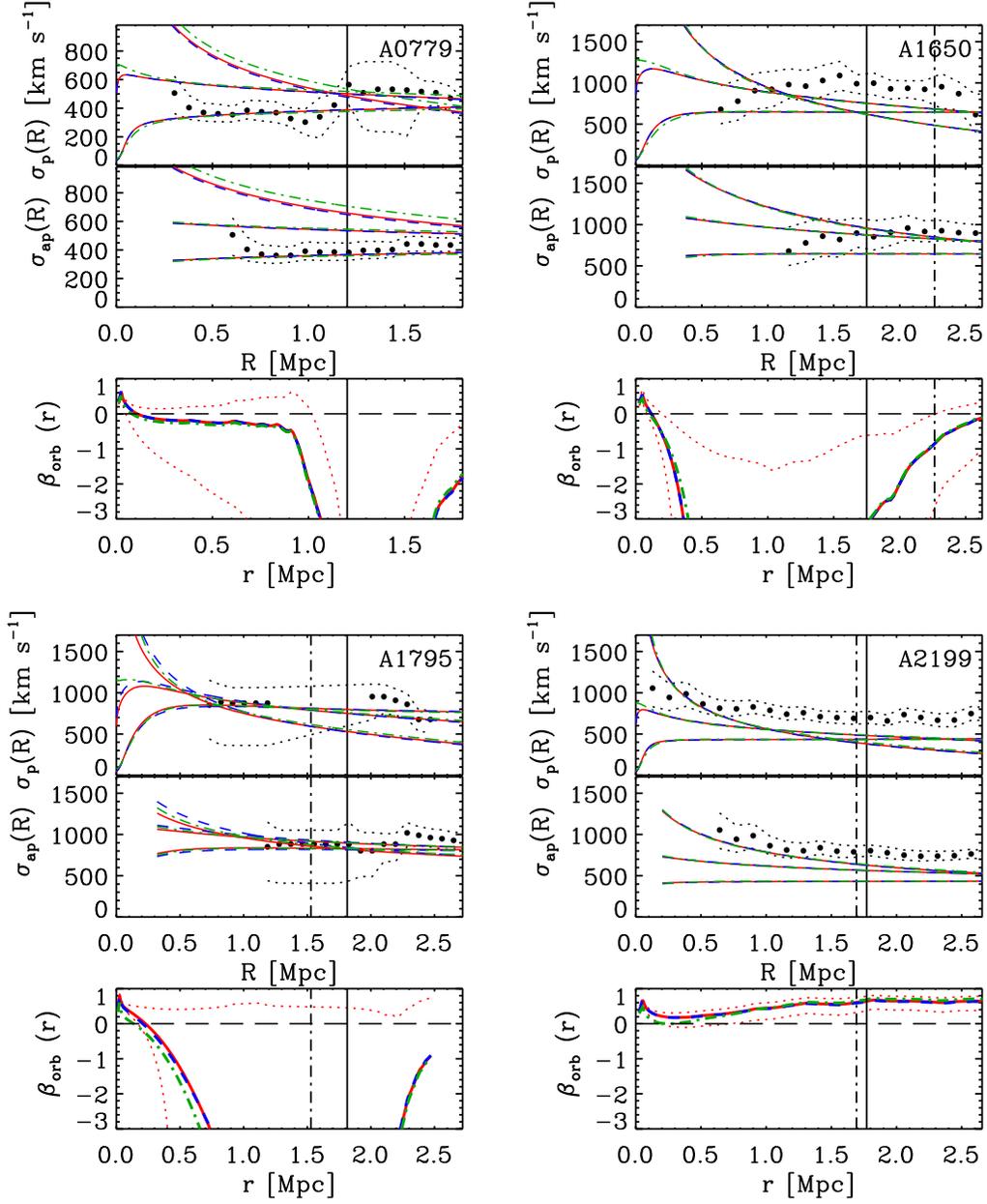}
\caption{Same as Fig. \ref{fig-orbit1}, but for late-type galaxies in
A779, 1650, 1795, and 2199.}\label{fig-orbitem}
\end{figure}

We present, in Figures \ref{fig-disp1} and \ref{fig-disp2},
  the VDPs for all, early-type, and late-type galaxies with $\delta\leq2.0$ 
  in each cluster. 
We computed the velocity dispersion of the
  galaxies lying within a bin with fixed radial width, 
  $\Delta R = R_{max}/15$ for all galaxies and $\Delta R = R_{max}/5$ 
  for the early-type and late-type galaxies as increasing the bin center by a fixed step width,
  $\delta R = R_{max}/40$. 
We stopped the calculation when the number of galaxies
  in a bin is less than 9, and used larger bin size if the number
of computed dispersions in a cluster is less than five.
We do not present the profiles for the subsample of the cluster whose final number
  of computed dispersions is less than five.
The velocity dispersions for our sample cluster are not similar,
but show diverse features. In addition, they are not always constant
throughout the radius.

\section{Galaxy Orbits in Clusters}\label{result}
If we assume spherical symmetry of a collisionless galaxy cluster, we
can apply the Jeans equation in the absence of rotation to the
dynamical analysis of the cluster. The spherical Jeans equation is

\begin{equation}
{d\over{dr}}\, \left( \nu_{\rm g}(r) \sigma_r^2(r) \right)
+ {{2\,\beta_{\rm orb}(r)}\over{r}}\, \nu_{\rm g}(r) \sigma_r^2(r)
= - \nu_{\rm g}(r)\,{{G M_{\rm tot}(r)}\over{r^2}}\ , \label{jeans}
\end{equation}

where
$\sigma_r(r)$ is the radial component of velocity dispersion,
  $\beta_{\rm orb}(r)\equiv 1-\sigma_\theta^2(r)/\sigma_r^2(r)$ is 
  the velocity anisotropy\footnote{It is noted that $\beta_{\rm orb}(r)$ is velocity anisotropy, 
  while $\beta$ is a parameter used for X-ray mass profile in \S \ref{xmass}.},
  $G$ is the gravitational constant,
  and $M_{\rm tot}(r)$ is the total gravitating mass contained
  within a sphere of radius $r$ (e.g. \citealt{bt87}). 
$\sigma_\theta(r)$ is the tangential component of velocity dispersion that is
  equal to the azimuthal component, $\sigma_\phi(r)$, 
  in the absence of cluster rotation. However, cluster rotation is not negligible
  in some clusters \citep{hl07}.

With an aid of independent determination of the cluster mass profile
using X-ray data,
we determine the velocity anisotropy (orbits) of cluster galaxies using two methods:
(1) from the comparison of the calculated VDP using the Jeans equation with the
measured VDP (e.g., \citealt{cote01,hwang07}), and
(2) to calculate directly $\beta_{\rm orb}(r)$
using the Jeans equation (e.g., \citealt{nk96,ben06}).

\subsection {Method 1}\label{method1}

Our strategy to determine the velocity anisotropy of cluster galaxies is as
follows: (1) With the galaxy number density profile, $\nu_{\rm g}(r)$,
of all, early-type, and late-type galaxies and the mass profile, $M_{\rm tot} (r)$, in hand,
assuming constant velocity anisotropy, $\beta_{\rm orb}$, over the radius in prior,
we compute the theoretical projected VDP,
$\sigma_p(R)$ and theoretical
projected aperture VDP, $\sigma_{ap}(\le R)$,
using the Jeans equation; (2) From the comparison of these calculated
VDPs with measured VDPs, we
determine the velocity anisotropy of cluster galaxies.

We begin by deriving the theoretical projected dispersion
profiles. The equation (\ref{jeans}), spherical Jeans equation can
be solved for the radial component of velocity dispersion,
$\sigma_r(r)$:

\begin{equation}
\sigma_r^2(r)
={1\over{\nu_{\rm g}(r)}}\,
\exp\left( -\int {{2\beta_{\rm orb}}\over{r}}\,dr \right)\,
\left[ \int_r^{\infty} \nu_{\rm g}\,{{GM_{\rm tot}}\over{x^2}}\,
\exp\left( \int {{2\beta_{\rm orb}}\over{x}}\,dx \right)\, dx \right]\ .
\label{eq5}
\end{equation}

Then the projected VDP, $\sigma_p(R)$ can
be derived by

\begin{equation}
\sigma_p^2(R)
= {2\over{\Sigma_{\rm g}(R)}}\,
\int_R^{\infty} \nu_{\rm g} \sigma_r^2(r)
\left(1 - \beta_{\rm orb}\,{{R^2}\over{r^2}} \right)\, {{r\,dr}\over{\sqrt{r^2 - R^2}}}, \label{eq-vdisp}
\end{equation}

where $R$ is the projected clustercentric distance, 
  and $\Sigma_{\rm g}(R)$ is the projected density profile
  that is a projection of the three-dimensional density profile $\nu_{\rm g}(r)$:

\begin{equation}
\Sigma_{\rm g}(R) = 2\int_{R}^{\infty} \nu_{\rm g}(r)
{{r\,dr}\over{\sqrt{r^2-R^2}}}\ . \label{eq7}
\end{equation}

The projected aperture dispersion profile, $\sigma_{ap}(\le R)$, which
is the velocity dispersion of all objects interior to a given
projected radius $R$, can be computed by

\begin{equation}
\sigma_{\rm ap}^2(\le R) =
  \left[ \int_{R_{\rm min}}^R \Sigma_{\rm g}(R^\prime) \sigma_p^2(R^\prime)\,
         R^\prime\,dR^\prime \right]\,
  \left[\int_{R_{\rm min}}^R \Sigma_{\rm g}(R^\prime)\,R^\prime\,dR^\prime
         \right]^{-1}\ ,
\label{eq8}
\end{equation}

where $R_{\rm min}$ is the projected clustercentric distance of the
innermost data point for the cluster galaxies.

For all galaxies in a cluster,
  we present the measured VDP compared with the
  calculated VDP for different velocity anisotropies
  in Figures \ref{fig-orbit1}, \ref{fig-orbit2}, and \ref{fig-orbit3}.
The top panels show the projected VDPs, and the middle panels show the projected
  aperture VDPs.
The measured dispersion data taken from
  Figures \ref{fig-disp1} and \ref{fig-disp2} are
  shown by filled circles along with their confidence intervals. 
The projected aperture dispersion profiles are
  plotted using a similar fashion to the case of the top panel.
Although it is difficult to distinguish the velocity anisotropy
  clearly in the top panel (the middle panels show more stable result),
  it appears that the galaxy orbits in clusters show diverse patterns.
Three results based on NFW, Hernquist, and KEK05 profiles of galaxy number density
  appear to be similar for a cluster.
In Figures \ref{fig-orbitab} and \ref{fig-orbitem},
  we show the results of similar analysis for early-type and late-type galaxies, respectively.
The difference between subsamples in a cluster is not
  clearly seen at this stage, and we discuss in detail the individual cluster in Appendix A.

\subsection {Method 2}\label{method2}

We combine the equation (\ref{jeans}), spherical Jeans equation
with the equation (\ref{eq-vdisp}) that defines
the projected VDP
in order to compute $\beta_{\rm orb}(r)$ as a function of the radius directly.

Replacing $\beta_{\rm orb}$ in equation (\ref{eq-vdisp}) with that in equation (\ref{jeans}),
we get,

\begin{eqnarray}
\frac{1}{2} \left[\,\, \Sigma_{g}(R) \ \sigma_{p}^{2}(R) - \,
{R^2}{\int_{R}^{\infty}}{\frac {G M_{\rm tot} (r) \nu_{g}}{r^{2} {\sqrt{{r^2} - {R^{2}}}}}} \,dr \right]
\nonumber \\ = \,\, \,\,
{\int_{R}^{\infty}}{\frac {r \nu_{g} \sigma_{r}^{2} dr}{{\sqrt{{r^2} - {R^{2}}}}}} +
{\frac {R^{2}}{2}}{\int_{R}^{\infty}}{\frac {d (\nu_{g} \sigma_{r}^{2})}{dr}}{\frac {dr}{{\sqrt{{r^2} - {R^{2}}}}}}.
\end{eqnarray}

This equation can be reduced by integrating the second term on the right-hand side by parts \citep{bm82},

\begin{eqnarray}
\frac{1}{2} \left[\Sigma_{g}(R) \sigma_{p}^{2}(R)
- {R^{2}{\int_{R}^{\infty}} \frac {G M_{\rm tot} (r) \nu_{g}}
{r^{2} {\sqrt{{r^2} - {R^{2}}}}}} dr \right]\nonumber \\
= \int_{R}^{\infty} \frac {(\frac {3 R^{2}}{2} - r^{2})}{{\sqrt{{r^2} - {R^{2}}}}}\frac { d ( \nu_{g} \sigma_{r}^{2})}{dr} \,dr.
\end{eqnarray}

Then, $\nu_{g} \sigma_{r}^{2}$ can be expressed as a sum of four integrals (see \citealt{bic89} for detail),

\begin{equation}
\nu_g(r) \sigma_r^2 = {\rm I_1(r)} - {\rm I_2}(r) + {\rm I_3}(r) - {\rm I_4}(r),
\end{equation}

\begin{equation}
{\rm I_1}(r) = \frac{1}{3} \int_r^{\infty} \frac{G M_{\rm tot}(r) \nu_g}{r^2} \ud r,
\end{equation}

\begin{equation}
{\rm I_2}(r) = \frac{2}{3r^3} \int_0^r G M_{\rm tot}(r) \nu_g r \ud r,
\end{equation}

\begin{equation}
{\rm I_3}(r) = \frac{1}{r^3} \int_0^r R \Sigma_g(R) \sigma_{p}^2(R) \ud R,
\end{equation}

\begin{eqnarray}
{\rm I_4}(r) = \frac{2}{\pi r^3} \int_r^{\infty} R \Sigma_g(R) \sigma_{p}^2(R)
\Big( \frac{r}{\sqrt{R^2-r^2}} - \sin^{-1} \frac{r}{R} \Big) \ud R.
\end{eqnarray}

After computing $\nu_{g} \sigma_{r}^{2}$, we finally obtain $\beta_{\rm orb}(r)$ from the
Jeans equation,

\be
\beta_{\rm orb}(r) = - \frac{r}{2 \nu_g \sigma_r^2}
\Big[ \frac {G M_{\rm tot}(r) \nu_g}{r^2} + \frac{\ud}{\ud r} (\nu_g \sigma_r^2)
\Big].
\ee

Practically, all integrations up to infinity were performed using a large radius, 
  $R_t=1.5R_{max}$ at which both $\nu_{\rm g} \sigma_{r}^{2}$ and $\Sigma_{\rm g}(R)$
  approach to zero. 
In addition, input VDPs derived in \S \ref{vdp} were smoothed for stable computation.
We compute $\beta_{\rm orb}$ using the above equation, and
  present the results in the bottom panels of Figures \ref{fig-orbit1}, \ref{fig-orbit2}, and \ref{fig-orbit3}.
The errors of $\beta_{\rm orb}$ are computed using the upper and lower confidence intervals
  of the VDPs shown in Figures \ref{fig-disp1} and \ref{fig-disp2}.
The behavior of $\beta_{\rm orb}$ computed in this Section is
  similar to that in \S \ref{method1}. 
It is noted that the calculated $\beta_{\rm orb}$ may not be reliable in the very inner region
  where the observed VDPs are not available (interpolation of the observed VDPs in the outer region is used)
  and X-ray mass profiles are not properly determined due to the low resolution of X-ray instrument.
Three results based on NFW, Hernquist, and KEK05 profiles of galaxy number density 
  are also similar for a cluster.
We also show the results of similar analysis for early-type and late-type galaxies
  in Figures~\ref{fig-orbitab} and \ref{fig-orbitem}, respectively.
Detailed discussion of the individual cluster is given in Appendix A.

\section{Global Cluster Properties}\label{global}
\subsection {Cluster Morphology}\label{morph}

In order to investigate the connection between cluster morphology and cluster dynamics,
it is useful to determine the ellipticity and orientation of a cluster
using the spatial distribution of the member galaxies.
To determine the cluster shape, we employ the dispersion ellipse of the
bivariate normal frequency function of position vectors (see, e.g., \citealt{trumpler53,carter80,bur04,hl07}).
The dispersion ellipse is defined by \citet{trumpler53} as the contour at which the density is 0.61 times the maximum density
of a set of points distributed normally with respect to two correlated variables, although
the points need not be distributed normally in order to determine the proper cluster shape.
From the first five moments of the spatial distribution,

\begin{mathletters}
\begin{equation}
\mu_{10} = \frac{1}{N}\sum_{i=1}^N X_i, ~~
\mu_{01} = \frac{1}{N}\sum_{i=1}^N Y_i
\end{equation}
\begin{equation}
\mu_{20} = \frac{1}{N}\sum_{i=1}^N X^2_i - \left( \frac{1}{N}\sum_{i=1}^N X_i \right)^2
\end{equation}
\begin{equation}
\mu_{11} = \frac{1}{N}\sum_{i=1}^N X_iY_i - \frac{1}{N^2}\sum_{i=1}^N X_i\sum_{i=1}^N Y_i
\end{equation}
\begin{equation}
\mu_{02} = \frac{1}{N}\sum_{i=1}^N Y^2_i - \left( \frac{1}{N}\sum_{i=1}^N Y_i \right)^2 \; \; \mbox{,}
\end{equation}
\end{mathletters}

where X and Y are clustercentric distances in the direction of right ascension
and declination, respectively, the semimajor and semiminor axes of the ellipse, $\Gamma_A$ and $\Gamma_B$,
are derived by solving the equation
\begin{equation}
\left| \begin{array}{cc}
\mu_{20}-\Gamma^2 & \mu_{11} \\
\mu_{11}     & \mu_{02}-\Gamma^2 \\
\end{array} \right|
= 0 \; \; \mbox{.}
\end{equation}

The position angle of the major axis, measured from north to east, is given by
\begin{equation}
\Theta_2 = \cot^{-1}\left( -\frac{\mu_{02} - \Gamma_A^2}{\mu_{11}}\right) + \frac{\pi}{2} \; \; \mbox{,}
\end{equation}
and the ellipticity is defined by
\begin{equation}
\epsilon = 1 - \frac{\Gamma_B}{\Gamma_A} \; \; \mbox{.}
\end{equation}

The major and minor axes, position angles, and ellipticities derived
for our sample clusters are listed in Table \ref{tab-mor},
and the dispersion ellipse is shown in Figure \ref{fig-spvel}.
The ellipticities are in the range $\epsilon=0.15-0.36$, indicating no strong elongation.

\subsection {Analysis of Substructure} \label{sub}
The analysis of substructure is a useful diagnostic tool for
understanding the dynamical state of galaxy clusters. A useful
discussion of several substructure tests is given in
\citet{pin96}. As described in \citet{hl07}, we derived number
density maps using the spatial position (two-dimensional; 2D) of
member galaxies, and performed one-dimensional (1D) and
three-dimensional (3D) substructure tests for our sample clusters.

The majority of 1D (velocity histogram) substructure tests are normality tests. 
We present the results of five 1D tests in Table \ref{tab-sub}. 
The values of the I-test \citep{tea90} are shown in
  the second and the third column; $I_{90}$ is the critical value
  for rejecting the Gaussian hypothesis at 90\% confidence.
Therefore, a velocity distribution is considered to be
  non-Gaussian if $I>I_{90}$. 
We find that four of the sample clusters 
  (A779, 1650, 1651, and 2199) do not satisfy the
  Gaussian hypothesis using the I test.

The skewness, which is a measure of the degree of asymmetry of a distribution, and the
confidence level at which it rejects normality are given in the fourth and the fifth columns of Table \ref{tab-sub}.
Positive or negative skewness indicates that the distribution is skewed to the right or left, respectively,
with a longer tail to one side of the distribution maximum.
The kurtosis, which is the degree to which a distribution is peaked, and the
confidence level at which it rejects normality are given in the sixth and the seventh columns.
Positive values indicate pointed or peaked distributions,
while negative values indicate flattened or non peaked distributions.
The skewness test rejects a Gaussian distribution with a confidence of over $90\%$ for A85, 1651, and 2199.
The kurtosis test rejects the hypothesis of Gaussianity for A779, 1650, 1651, 2034, and 2199
with confidence of over $90\%$.

From the eighth to the eleventh column of Table \ref{tab-sub},
we present the asymmetry index (AI) and the tail index (TI)
introduced by \citet{bb93} along with their confidence levels.
The AI measures the symmetry in a population by comparing gaps in the data on the
left and right sides of the sample median, and TI compares the spread of the data at the $90\%$
level with the spread at the $75\%$ level.
The Gaussian hypothesis for A779, 1800, and 2199 are rejected by the AI test,
and that for A85, 1650, 2199 by the TI test with a confidence level of over $90\%$.

\begin{figure}
\includegraphics [width=125mm]{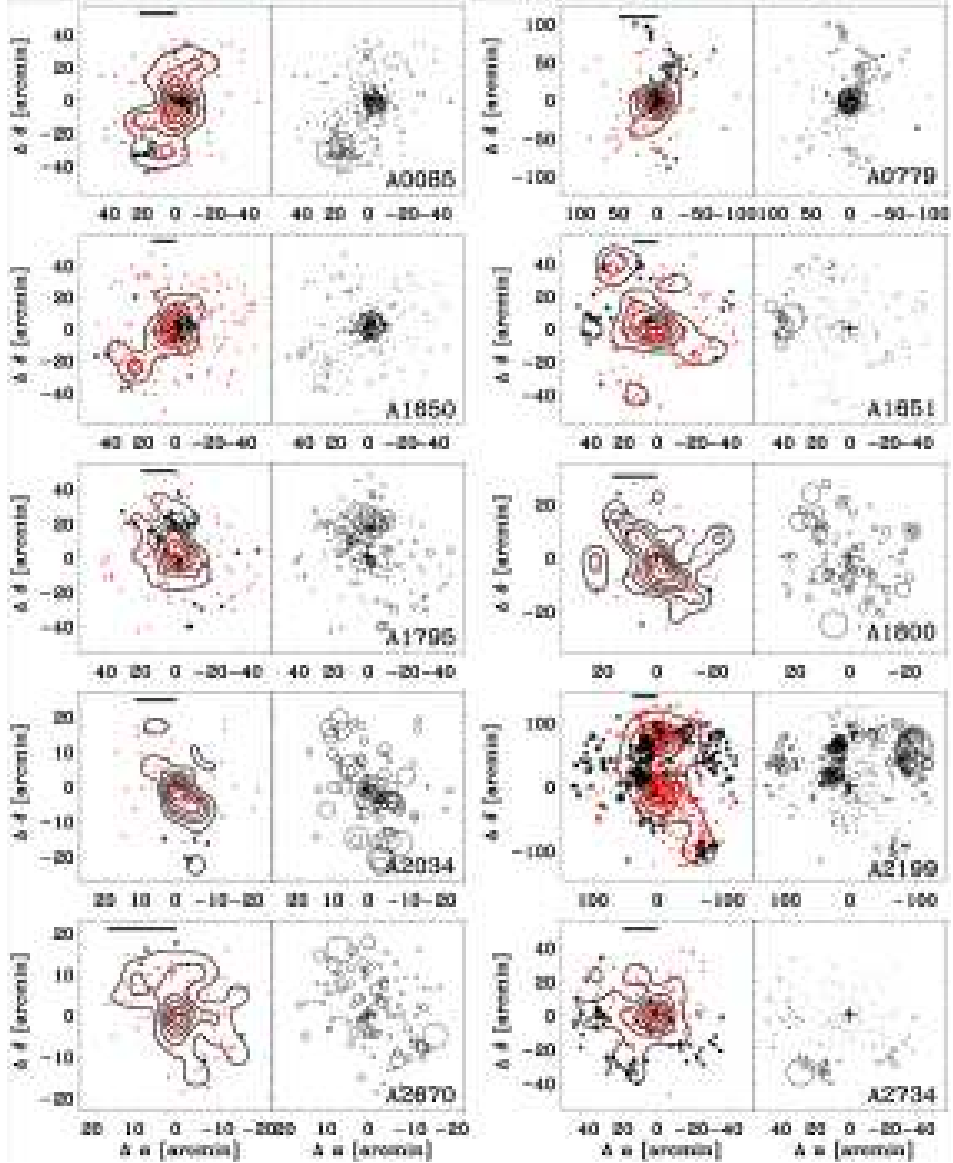}
\centering
\caption{{\it Left panels}: 
Galaxy number density contour maps for our sample clusters. 
The member galaxies with $\delta\leq2.0$ and $\delta>2.0$ are represented by dots and crosses, respectively.
The number density contours are overlaid. The plus signs indicate the
cluster centroids, and the thick horizontal bars represent a
physical extent of 1 Mpc. {\it Right panels}: Dressler-Shectman
plots for the same clusters. Each galaxy is plotted by a circle
with diameter proportional to $e^\delta$. North is up, and east is
to the left.}\label{fig-sub}
\end{figure}

We have constructed number density contours for the clusters
  using different bin sizes of $0.25 R' \times 0.25 R' ~\rm{Mpc^2}$ depending on
  the cluster size [$R'=4(\Gamma_A \Gamma_B)^{1/2}/1.5$].
$\Gamma_A$ and $\Gamma_B$ are in units of Mpc, 4 is an arbitrary constant,
  and 1.5 (Mpc) is a normalization constant. 
We have smoothed the contours
  using a cubic convolution interpolation method. The contour interval,
  (max density in cluster)$/6$, is also
  determined according to the maximum number density of the clusters.
We plot the number density contour map in the first and third columns of Figure \ref{fig-sub}.

Using the velocity data and positional information on the
galaxies, we have performed a $\Delta$-test \citep{ds88}, which
computes local deviations from the systemic velocity ($v_{\rm
sys}$) and dispersion ($\sigma_p$) of the entire cluster. For each
galaxy, the deviation is defined by

\begin{equation}
\delta^2 =  \frac{N_{nn}}{\sigma_p^2} \left[ (v_{\rm local}-v_{\rm sys})^2 +
(\sigma_{\rm local}-\sigma_p)^2 \right] \; \; \mbox{,}
\end{equation}

where $N_{nn}$ is the number of galaxies that defines the local environment, taken
to be $\sim {N_{gal}}^{1/2}$ in this study. The sum of $\delta$ over all galaxies
in a cluster, $\Delta$,
is used to quantify the presence of substructure.
It is approximately equal to the total number
of galaxies in a cluster in the case of no substructure, while it is larger
than the total number of galaxies in a cluster in the presence of substructure.

The statistical significance of the deviation is computed by Monte Carlo
simulations. Velocities are randomly
assigned to the galaxies at their observed positions, and $\Delta_{\rm sim}$ is
computed for each simulated cluster. We construct 1000 simulated clusters and
compute $\Delta_{\rm sim}$ for each simulation. We present $\Delta_{\rm obs}$
which is computed using real data and the fraction of simulated clusters with
$\Delta_{\rm sim}>\Delta_{\rm obs}$ in the final two columns of Table \ref{tab-sub}.
Small values of $f(\Delta_{\rm sim}>\Delta_{\rm obs})$ indicate statistically
  significant substructure.
A779, 1795, 2199, and 2734 have much smaller values than other clusters,
  indicating significant substructures.

We plot the positions of cluster galaxies, represented by circles with
  radii proportional to $e^{\delta}$, in the second and the fourth columns of Figure \ref{fig-sub}.
A large circle denotes a galaxy that is deviant in either velocity or
  dispersion compared with nearby galaxies; therefore, groups of large circles indicate
  the presence of substructure. 
No strong substructures in A85, 1650, 1651, 1800, 2034, and 2670 were found
  with the $f(\Delta_{\rm sim}>\Delta_{\rm obs})$ test, as confirmed in Figure \ref{fig-sub}.
In the number density contour maps of Figure \ref{fig-sub},
  we present the galaxies with $\delta\leq2.0$ that were used for orbit analysis ({\it dots})
  and those with $\delta>2.0$ in the substructure ({\it crosses}) separately,
  in order to show the usefulness of $\delta$ parameter to identify the substructure.

\subsection{Dynamical Status of the Clusters}\label{status}

\begin{figure}
\centering
\includegraphics [width=135mm]{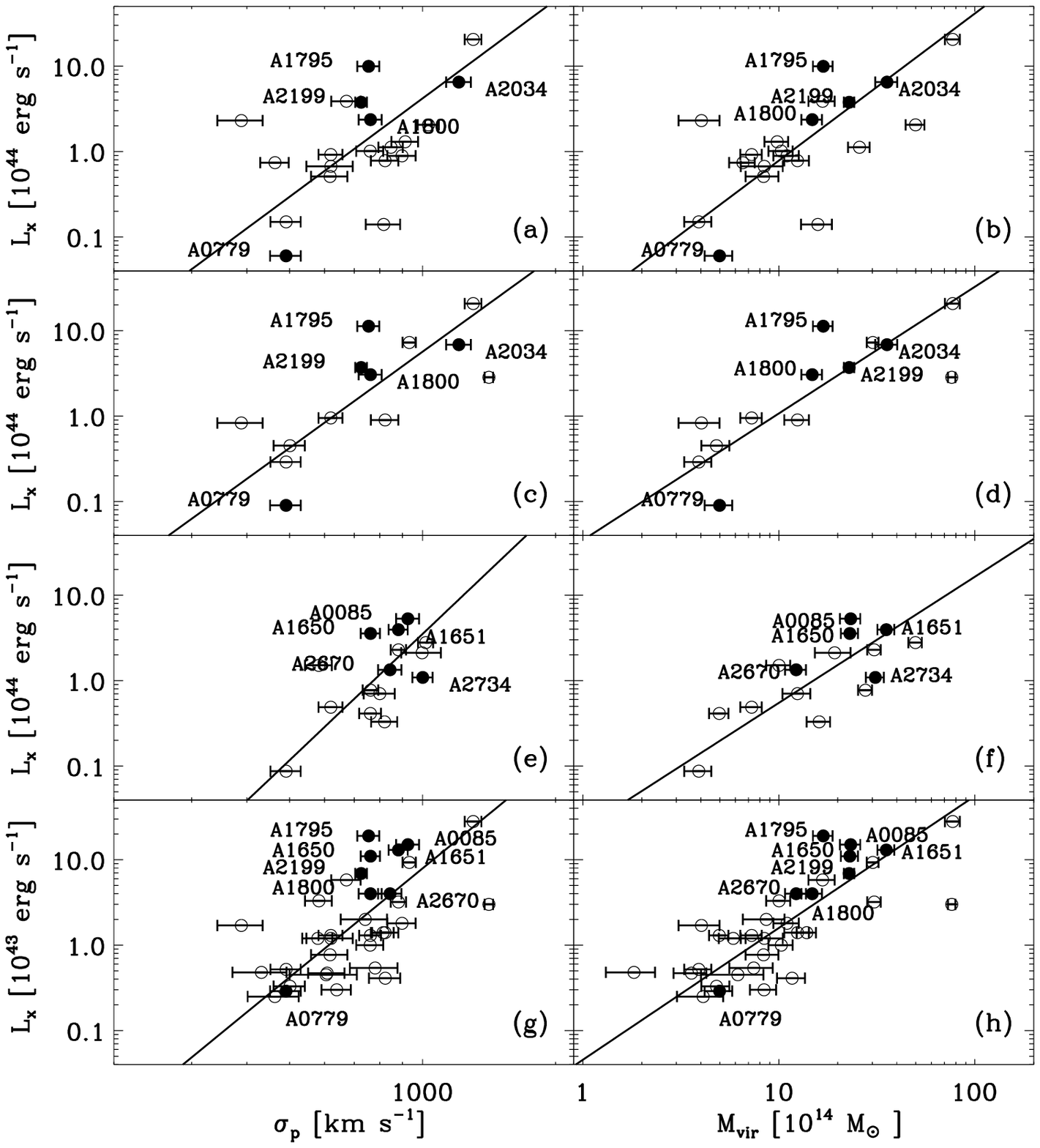}
\caption{X-ray luminosity $L_{\rm x}$ (0.1$-$2.4 keV) vs. the
velocity dispersion and virial mass for the our sample
clusters (filled circles) compared with the other clusters (open circles)
for the 113 selected clusters.
X-ray luminosities in (a) and (b) are from \citet{boh00},
those in (c) and (d) from \citet{ebe98,ebe00}, those in
(e) and (f) from \citet{boh04}, and those (0.5$-$2.0 keV) in
(g) and (h) from \citet{ledlow03}. The solid lines indicate the
best fit for each panel. }\label{fig-xray}
\end{figure}

A study of the dynamical state for our sample clusters is useful for interpreting the galaxy orbits in clusters.
It is expected from self-similar models that a relationship $L_{\rm x}\propto \sigma_p^4$
  between the X-ray luminosity and the velocity dispersion of the clusters will exist (e.g. \citealt{qm82}).
For galaxy clusters in SDSS, \citet{popesso05} found $L_{\rm x}\propto \sigma_p^{3.68\pm0.25}$.
For galaxy clusters in 2dFGRS, \citet{hilton05} found
  a relation $L_{\rm x}\propto \sigma_p^{4.8\pm0.7}$ and
  suggested that high-$L_{\rm x}$ clusters are more dynamically evolved systems than the low-$L_{\rm x}$ clusters.
In Figure \ref{fig-xray},
we plot the X-ray luminosity as a function of velocity dispersion and virial mass
  for our sample clusters ({\it filled circles})
  for which X-ray luminosities are available in the literature,
  compared with other clusters ({\it open circles}) out of the 113 selected clusters.
We plot the X-ray luminosities from different literature sources separately.
It shows that $L_{\rm x}$ correlates well with $\sigma_p$ and $M_{\rm vir}$.
The power-law slopes are found to be in the range $3.7-4.9$ depending on X-ray references,
  which is consistent with the results of previous studies.
Our sample clusters except for A779 are usually X-ray bright compared with the other clusters,
  because it is easy to derive X-ray mass profiles for the X-ray bright clusters.
Interestingly, A779 and A1795 show a significant deviation from the relation 
$L_{\rm x}$-$\sigma_p$ and $L_{\rm x}$-$M_{\rm vir}$.

\begin{figure}
\centering
\includegraphics [width=135mm]{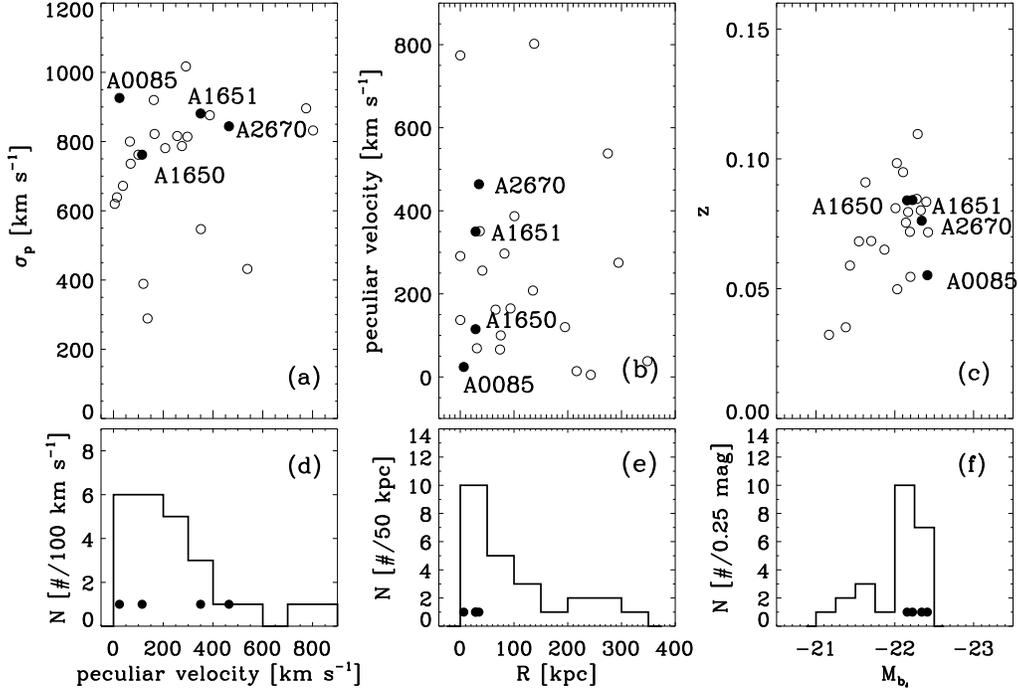}
\caption{(a) Velocity dispersion as a function of the absolute value
  of the peculiar velocities of the BCGs,
(b) absolute value of peculiar velocity as a function of the clustercentric
  distances of the BCGs, and
(c) redshift as a function of the absolute magnitudes of the BCGs
  in the $b_{\rm J}$ band for our sample clusters (filled circles)
  compared with the other clusters (open circles) among the 113 selected
  galaxy clusters. 
Histograms of the peculiar velocities (d),
the clustercentric distances (e), and the absolute magnitudes in
the $b_{\rm J}$ band (f) of the BCGs are also
shown for the clusters including our sample clusters.
The clusters in our sample are marked by
filled circles in (d), (e), and (f). }\label{fig-pec}
\end{figure}

The dynamics of the brightest cluster galaxies (BCGs) or cD galaxies in galaxy clusters
are also useful for understanding the
formation history of the clusters (see, e.g., \citealt{oh01}).
In particular, the peculiar velocity of the BCG, defined
by $v_p = v_{BCG}-v_{\rm cl}$, where $v_{BCG}$ is the observed velocity of BCGs
and $v_{\rm cl}$ is the mean velocity of the cluster,
is a useful indicator of the dynamical state of a cluster \citep{oh01}.
To estimate the peculiar velocities of BCGs in clusters,
we first identified the BCGs that have the brightest
$b_{\rm J}$ magnitudes in the catalog of member galaxies for the 113 selected galaxy clusters.
Then we conducted a visual inspection of cluster images
to determine whether there are any galaxies brighter than the selected BCGs,
using the catalog of member galaxies.
Since some very bright galaxies in clusters were not covered
as a result of observational difficulties such as fiber collision and saturation,
we finally selected 24 galaxy clusters for which the BCGs in the catalog agree with those in the images.
Using this sample of 24 galaxy clusters, we plot, in Figure \ref{fig-pec},
the cluster velocity dispersion as a function of the absolute value of the BCG peculiar velocity,
the absolute value of the BCG peculiar velocity as a function of clustercentric distance,
and the redshift of the clusters as a function of the BCG absolute magnitude in the $b_{\rm J}$ band.

It appears that the absolute values of the peculiar velocities of the BCGs are in the range 5$-$802 km s$^{-1}$
and that the median value of the peculiar velocities is 165 km s$^{-1}$.
The mean uncertainty on the values of the peculiar velocities is 117 km s$^{-1}$.
In addition, the clustercentric distances of the BCGs are in the range 0$-$350 kpc,
and the median value of the clustercentric distances is 75 kpc.
Interestingly, the BCG peculiar velocities for two (A1651 and A2670) clusters
are larger than that median value, indicating dynamical non-equilibrium.

\section{Discussion}\label{discuss}

We summarize the global kinematic properties for our sample clusters in Table \ref{tab-sum}.
The first column gives the Abell identification. The second through fourth columns list
  the existence of substructure as indicated by the 1D, 2D, and 3D tests, respectively :
``Yes'' is assigned 
  if three of five tests indicate the presence of substructure for 1D test,
  if the substructure at the contour level of 2$\times$(max density in cluster)/6 is seen for 2D test, and
  if f($\Delta_{sim}>\Delta_{obs}$)$<0.1$ for 3D test.
The cluster morphology determined in \S \ref{morph}
  is given in the fifth column: ``spherical'' for ellipticity less than $0.2$ and ``elongated''
  for ellipticity greater than $0.2$. 
``?'' is given for A85 since the spatial coverage is incomplete.
The dynamical status determined in \S \ref{status} is given in the sixth and seventh columns.
For the scaling relation between the X-ray luminosity and the velocity dispersion or the virial mass,
``No'' is assigned if the deviation of X-ray luminosities from the scaling relations derived in this study 
  is larger than one standard deviation in at least four out of eight panels in Figure \ref{fig-xray}.
For the peculiar velocities of the BCGs,
``No'' is assigned if the significance of the peculiar velocity of the BCG,
   $S$ [$\equiv|v_p|/(\epsilon_{\rm BCG}+\epsilon_{\rm cl}^2)^{1/2}$], is larger than 3.
$\epsilon_{\rm BCG}$ is the measurement error of the BCG velocity
  and $\epsilon_{\rm cl}$ is defined by $\epsilon_{\rm cl}\equiv\sigma_{\rm cl}^2/N_{gal}$,
  where $\sigma_{\rm cl}$ is the velocity dispersion of the cluster.
The description on the result for the individual cluster is given Appendix A.

\subsection{Effects of Different X-ray Mass Profiles}
\begin{figure}
\centering
\includegraphics [width=135mm]{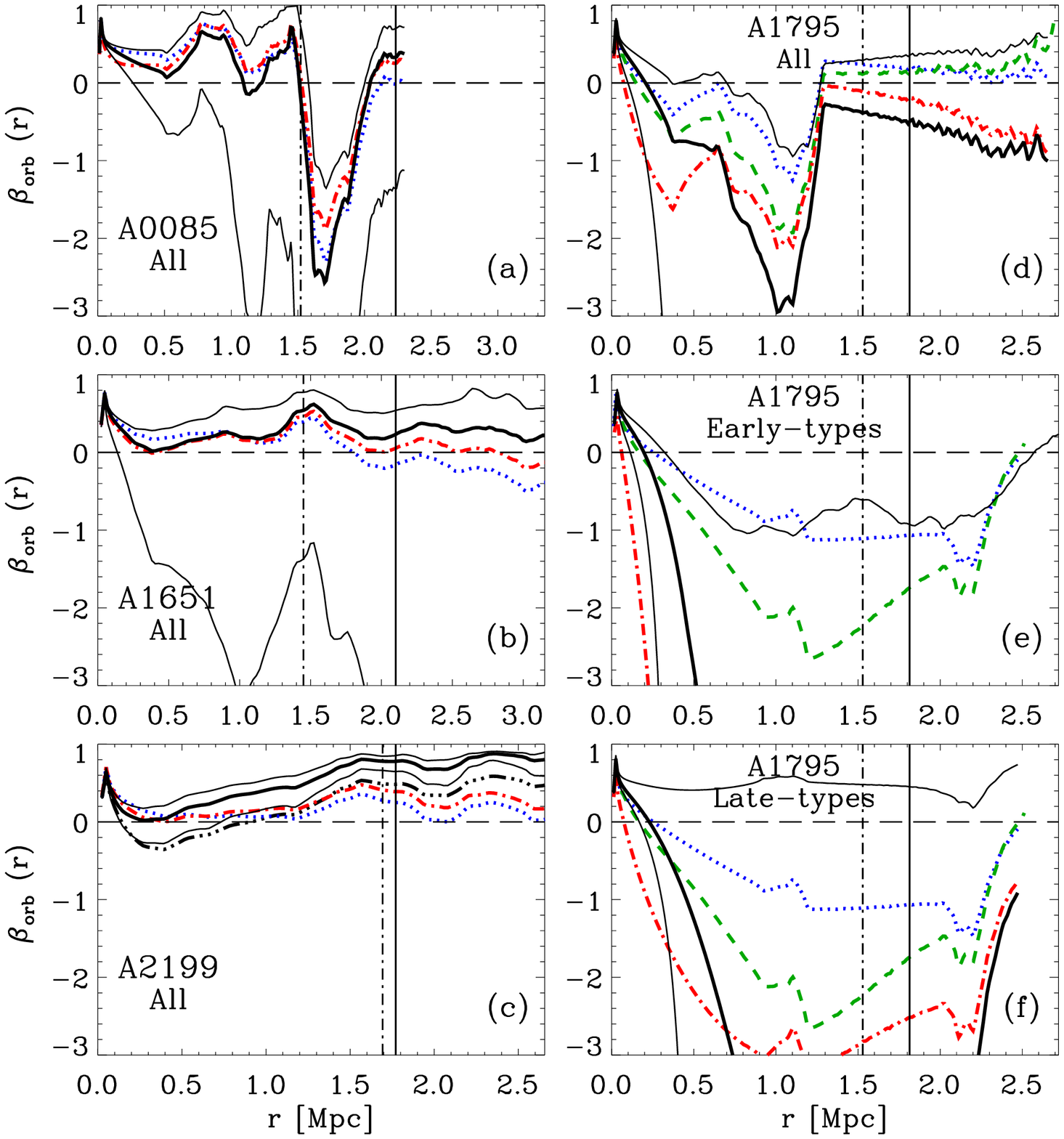}
\caption{VAPs for all galaxies of A85 (a), A1651 (b), A2199 (c),
  and those for all (d), early-type (e), late-type galaxies (f)
  of A1795 determined in \S \ref{method2} using the NFW profile,
  but based on several mass profiles:
 \citet[thick solid]{san03}, \citet[thick dotted]{dem03}, 
 \citet[thick dot-dashed]{rb02}, 
 \citet[thick dot-dot-dashed]{mar99}, and \citet[thick dashed]{vik06}.
The errors of VAPs using the mass profile of \citet{san03} are shown by thin solid lines.
The vertical dot-dashed and solid lines indicate the outer significance radius, $r_X$,
  in \citet{rb02} and $r_{200}$ radius computed in this study, respectively.
}\label{fig-masscomp}
\end{figure}

Since we used one X-ray mass profile per one cluster among several X-ray mass profiles,
  it is important to examine the results using different mass profiles.
Moreover, the mass profile of \citet{rb02} that we used in \S \ref{xmass},
  adopted a constant gas temperature model to derive X-ray mass profile,
  but recent X-ray observations showed that gas temperature
  is not constant over the clustercentric distance \citep{mar98,dm02,pra07}.
The isothermal model is known to
  make the mass profile steeper than that of non-isothermal models,
  leading to underestimate the cluster mass at small radii
  and overestimate at large radii.
\citet{mar98} showed that the isothermal model leads to
  underestimate the cluster mass by a factor of 0.74 of the non-isothermal model at one core radius
  and overestimate by a factor of 1.43 of the non-isothermal model at six core radius.
To investigate the effects of different mass profiles on the determination of galaxy orbits,
  we compute $\beta_{\rm orb}$ in each cluster with various mass profiles, and
  present the results in Figure \ref{fig-masscomp}.
The overestimation of the cluster mass from the isothermal model of \citet{rb02}
  is expected to give larger value of calculated VDP than the true value,
  and the value of $\beta_{\rm orb}$ would decrease.
Indeed it is seen that the cluster masses of the isothermal model in \citet{rb02}
  are larger than those of non-isothermal model in \citet{san03} at $r\sim r_{200}$ for A1651 and A2199
  (Fig. \ref{fig-mass}).
Therefore, the calculated $\beta_{\rm orb}$ with the isothermal model in \citet{rb02}
  are smaller than those with the non-isothermal model \citet{san03} [Fig. \ref{fig-masscomp} (b) and (c)].
However, the orbit determination does not significantly change,
  because these values of $\beta_{\rm orb}$ with various mass profiles are not much different 
  and accord within the uncertainty.
It is noted that the galaxy orbits can be determined differently
  as seen in the case of A2199 (isotropic orbit for \citealt{san03}, but tangential orbit for \citealt{mar99} 
  at $r\sim0.35$ Mpc; see Appendix \ref{a2199}).

The effect of the most accurate mass profile obtained with {\it Chandra} is seen in (d), (e), and (f).
Since the {\it Chandra} mass profile is available for only A1795 and the discrepancy
  among the mass profiles in A1795 is not large,
  the calculated $\beta_{\rm orb}$ with {\it Chandra} mass profile appears to
  accord with those from other mass profiles within the uncertainty.
The discrepancy becomes larger at large radius ($r>r_X$),
  but the reliability at this region is low 
  because the mass profile is determined from the extrapolation of that for the inner region.

\subsection{Effects of Different Galaxy Samples}\label{galsample}
Since significant fractions ($40-70\%$) of the clusters show substructures,
  indicating that they are in the process of merging \citep{jf99,ram07},
  it is important to study the effect of the presence of substructure
  on the determination of galaxy orbits.
It is expected that the presence of substructure leads to 
  an increase of the observed VDP and 
  the change of the galaxy number density profile.
As a result, the orbit determination can be different 
  depending on the degree of inclusion of the galaxies in the substructure.
Previously, \citet{bk04} showed that the galaxies in the substructure
  appear to have tangential orbit as a whole.
Therefore, the orbit analysis including the galaxies in the substructure
  may make $\beta_{\rm orb}$ have low values.

\begin{figure}
\centering
\includegraphics [width=135mm]{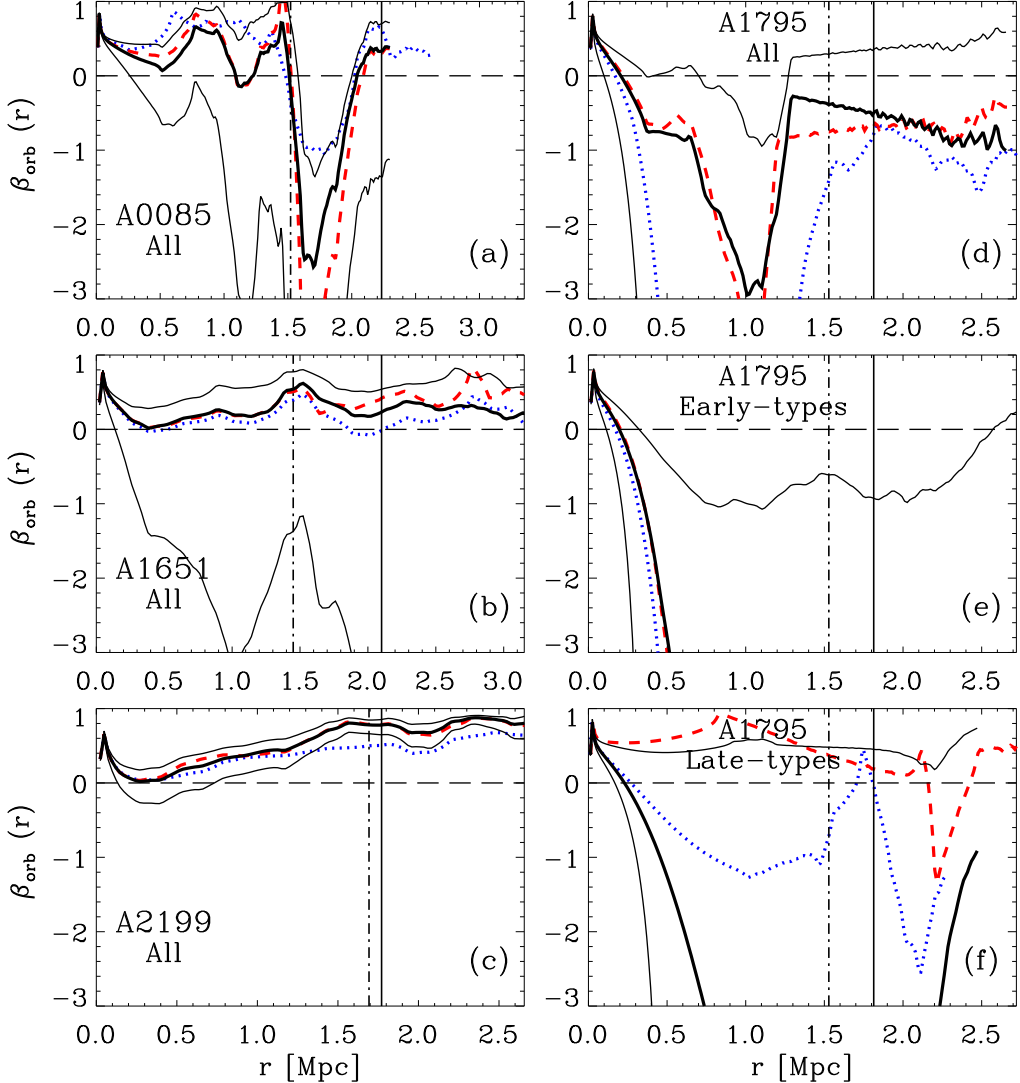}
\caption{VAPs for all galaxies of A85 (a), A1651 (b), A2199 (c),
  and those for all (d), early-type (e), late-type galaxies (f)
  of A1795 determined in \S \ref{method2} using the NFW profile,
  but based on several galaxy samples:
 galaxies with $\delta\leq2.0$ (thick solid),
 galaxies with $\delta\leq1.8$ (thick dashed), and
 galaxies with $\delta\geq0$ (thick dotted).
The errors of VAPs based on the galaxies with $\delta\leq2.0$ are shown by thin solid lines.
The vertical dot-dashed and solid lines indicate the outer significance radius, $r_X$,
  in \citet{rb02} and $r_{200}$ radius computed in this study, respectively.
}\label{fig-masscomp2}
\end{figure}

Figure \ref{fig-masscomp2} shows the effect of the presence of substructure on
  the determination of galaxy orbits.
As seen in (b)$-$(e),
  the calculated $\beta_{\rm orb}$ including the galaxies in the substructure ({\it dotted lines})
  tends to be lower values than those excluding the galaxies in the substructure ({\it solid and dashed lines}),
  being compatible with the results of \citet{bk04}.
However, A85 [Fig. \ref{fig-masscomp2} (a)] and late-type galaxies in A1795 [Fig. \ref{fig-masscomp2} (f)] 
  do not show similar behavior to other cases,
  since the changes in observed VDPs and in galaxy number density profiles are not similar to other clusters.
In conclusion, the orbit determination can be changed significantly [e.g., Fig \ref{fig-masscomp2} (f)],
  depending on the degree and the location of subclustering,
  but is not much changed for our sample clusters.



\subsection{Comparison with the Previous Studies}
For 10 Abell clusters,
  we have found that the orbits of galaxies 
   are consistent with isotropic orbits in most clusters,
   although it is difficult to conclude strongly for some clusters
   due to the large errors (e.g., A779, A1650) 
   and the variation as a function of clustercentric distance (e.g., A1795, A1800, A1795, A2199)
   in calculated $\beta_{\rm orb}$.
Isotropic galaxy orbits for majority of our sample clusters, 
  are consistent with
  those for composite clusters in the previous large cluster surveys
  (e.g., \citealt{car97a,car97b,van00,rines03}).
Existence of anisotropic galaxy orbits is also reported
  in other studies. For A2218, \citet{nk96}
  found that the galaxy orbits are tangential at small radii
  ($R\leq400$ kpc) and are radial at large radii. 
\citet{ben06} analyzed the galaxy orbits for five clusters up to $r\sim r_{200}$,
  and reported that A2199 and A496 are consistent with tangential orbits, 
  while A2390 radial orbits. Two clusters (MS1358 and A576)
  appear to have radial orbits, but these clusters may not be in
  hydrostatic equilibrium.

Numerical simulations showed that cluster
  galaxies are likely to follow isotropic orbits as seen in many
  clusters (e.g., \citealt{ghi98}). 
In detail, isotropic orbits are
  preferred in the inner region, and radial orbits in the outer region, 
  indicating existence of accreting galaxies from the cluster outskirt 
  \citep{crone94,cl96,tor97,tho98,ckk00,fal05}. 
However, tangential galaxy orbits found in the observational data, are not
  commonly seen in simulation data \citep{tor97,tho98}.
Interestingly, \citet{ben06} reported an offset in orbital profile
  for their observed and simulated clusters in the sense that the
  simulated clusters show preferentially tangential orbits.

\begin{figure}
\centering
\includegraphics [width=135mm]{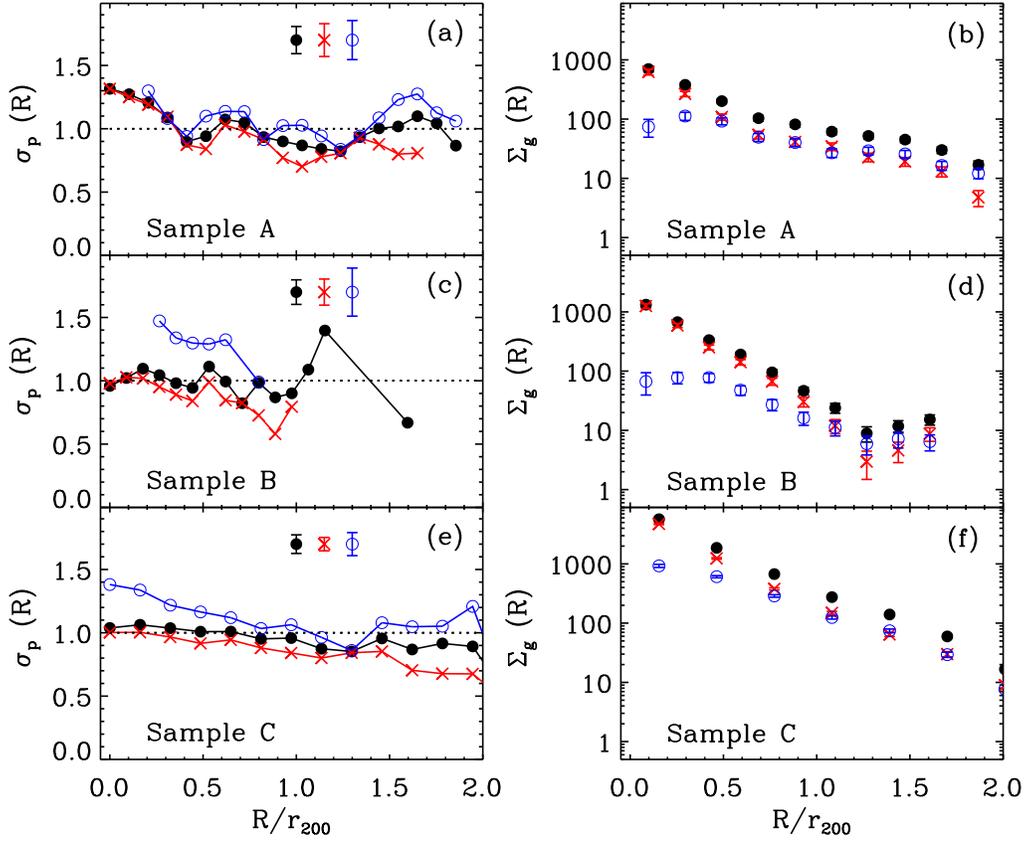}
\caption{(a) VDPs and (b) projected galaxy number density profiles for the clusters in Sample A; 
(c,d) those in Sample B; and (e,f) those in Sample C.
Filled circles, crosses, and open circles denote all, early-type, and
late-type galaxies, respectively. Typical errors of dispersion
profiles are shown by errorbars according to the subsamples.
}\label{fig-comp}
\end{figure}

For the orbital difference between early-type and late-type galaxies,
  the previous studies using the composite cluster, led to the conclusion
  that early-types have quasi-isotropic orbits and late-types radial orbits
  \citep{biv97,mah99,bk04}. 
Interestingly, the clusters in which orbital difference 
  among subsamples were studied in this study,
  show no significant difference between them.
In order to investigate the origin of difference between this study
  and the previous studies, 
  we show VDPs and projected galaxy number density profiles
  for several composite clusters in Figure \ref{fig-comp}. 
We construct the composite clusters for three different samples: 
  Sample A - using four clusters in which orbital properties for subsamples 
    are studied (A779, 1650, 1795, and 2199), 
  Sample B - using six clusters in which orbital properties among subsamples 
    are not studied (A85, 1651, 1800, 2034, 2670, and 2734), 
  and Sample C - using 62 clusters without definite merging evidence, selected in \S \ref{mem}.
All samples are constructed using the galaxies outside the substructure ($\delta\leq2.0$).

The VDP for the composite cluster is derived
  in the combined distribution of velocity relative to the cluster
  mean and normalized by the velocity dispersion of each cluster.
It is known that the shape of VDP provides some hints for the galaxy
  orbits in clusters, and varies depending on clusters
  \citep{dHK96,mah99,mur02,agu07}: flat, decaying, or rising VDP. 
It appears that the differences in VDPs and in projected galaxy number density profiles
   between early-type and late-type galaxies for Sample A
   are not significant at $0.3<R/r_{200}<0.8$ where the VDPs are available, 
   while those in Sample B are large.
In Sample C, the difference is significant with small errors: 
  the values of velocity dispersion for late-types are larger than those for early-types, 
  and the VDP for late-types is decreasing along the
  radius while that for early-types is nearly flat. 
Sample C is comparable to the composite cluster based on ENACS data used in
  \citet{bk04}, and our results for Sample C are similar to theirs.
Thus, if we adopt the result of \citet{kbm04} that
  early-types have isotropic orbits, then the orbits for late-types
  in our composite cluster (Sample C) can be regarded as radial orbits, 
  which is consistent with the results of the previous studies. 
Similarly, since not only VDPs but also projected galaxy number density profiles
  are not distinguishable in Sample A at $0.3<R/r_{200}<0.8$,
  it is expected that the orbital difference
  between early-type and late-type galaxies in Sample A is not significant.
If we use the galaxies with lower probability of belonging to the substructure (e.g., $\delta\leq1.4$),
  the results change little, indicating the effects of the galaxies substructure
  is not significant.
Therefore, no orbital difference between early-type and late-type galaxies in Sample A
  can be interpreted as their own characteristics of individual clusters.
Previously, \citet{rines05} also reported that the VDP of ELG is not
  different from the non-ELG for CAIRNS clusters, 
  but their spatial distributions are different.
In numerical simulations,
  \citet{dia01} found that there is no significant difference in
  orbital properties between red and blue galaxies : both red and
  blue galaxies are close to being isotropic. 
Thus, it is necessary to study what makes the diversity of galaxy orbits
  using more observational and simulation results.

\section{Summary}\label{sum}

We present the results of a study for the galaxy orbits in galaxy
clusters using a spectroscopic sample of galaxies in SDSS and
2dFGRS. Our results are summarized as follows:

\begin{enumerate}

\item We have determined the member galaxies of 731 and 230 Abell clusters
  covered in SDSS and 2dFGRS, respectively.
We have selected 10 clusters using three criteria :
  the number of member galaxies is greater than or equal to 40, 
  the spatial coverage is complete, 
  and X-ray mass profile is available in the literature.

\item For the selected 10 clusters,
  we derived the radial profile of the galaxy number density and
  velocity dispersion for all, early-type, and late-type galaxies 
  outside the substructure ($\delta\leq2.0$).

\item We have investigated the galaxy orbits for our sample
  clusters with constant and variable $\beta_{\rm orb}$ through the
  clustercentric radius using Jeans equation. 
The resulting galaxy orbits based on two methods appear to be consistent.

\item Using all member galaxies, 
  most of our sample clusters are found to be have isotropic orbits, 
  although it is difficult to conclude strongly for some clusters
  due to the large errors (e.g., A779, A1650)
  and the variation as a function of clustercentric distance (e.g., A1795, A1800, A1795, A2199)
  in calculated $\beta_{\rm orb}$.
For four clusters (A779, 1650, 1795, and 2199) 
  the orbital difference between early-type and late-type galaxies 
  appears not to be significant.

\item We have determined the cluster morphology
  using the dispersion-ellipse method, 
  and have found the ellipticity in the range of $0.15-0.36$, 
  indicating no strong elongation for our sample clusters.

\item We have investigated dynamical status for our sample
clusters using substructure (1D, 2D, and 3D) tests, the relation
between X-ray luminosity and the velocity dispersion or the virial mass,
and peculiar velocity of BCGs. It is found that the majority of
our sample clusters are dynamically relaxed system.

\end{enumerate}

\acknowledgments
We thank the referee, A. Biviano, for useful comments that improved
  significantly the original manuscript.
We also thank Woong-Tae Kim for useful comments on the galaxy number density profiles
  of clusters.
We thank A.J.R. Sanderson and M. Markevitch for kindly providing us the central gas
 temperature data of our sample clusters.
We also would like to thank all the people involved in creating
the SDSS, 2dFGRS and NED. Funding for the SDSS and SDSS-II has
been provided by the Alfred P. Sloan Foundation, the Participating
Institutions, the National Science Foundation, the U.S. Department
of Energy, the National Aeronautics and Space Administration, the
Japanese Monbukagakusho, the Max Planck Society, and the Higher
Education Funding Council for England. The SDSS Web Site is
http://www.sdss.org. The SDSS is managed by the Astrophysical
Research Consortium for the Participating Institutions. The
Participating Institutions are the American Museum of Natural
History, Astrophysical Institute Potsdam, the University of Basel,
the University of Cambridge, Case Western Reserve University, the
University of Chicago, Drexel University, Fermilab, the Institute
for Advanced Study, the Japan Participation Group, Johns Hopkins
University, the Joint Institute for Nuclear Astrophysics, the
Kavli Institute for Particle Astrophysics and Cosmology, the
Korean Scientist Group, the Chinese Academy of Sciences (LAMOST),
Los Alamos National Laboratory, the Max Planck Institute for
Astronomy, the Max Planck Institute for Astrophysics, New Mexico
State University, Ohio State University, the University of
Pittsburgh, the University of Portsmouth, Princeton University,
the US Naval Observatory, and the University of Washington. This
research has made use of the NASA/IPAC Extragalactic Database
(NED) which is operated by the Jet Propulsion Laboratory,
California Institute of Technology, under contract with the
National Aeronautics and Space Administration.
This work was supported in part by grant R01-2004-000-10490-0 and
R01-2007-000-20336-0 from the
Basic Research Program of the Korea Science and Engineering Foundation.

\appendix

\section{Properties of Individual Clusters}

\subsection{Abell 85}
A85 was covered in SDSS, but the outer region ($R>26\arcmin$) was
  not surveyed (see Fig. \ref{fig-spvel}). 
This cluster is in the complex of clusters Abell 85/87/89, and the member galaxies
selected in this study include only Abell 85/87. The structure of
this cluster is well studied by several authors
\citep{lima02,dur98,kem02,dur05}: a southern blob (or group) that
is merging from south ($\sim10\arcmin$ from the cluster center), a
subcluster that is merging from southwest ($\sim4\arcmin$ from the
cluster center), an extended 4 Mpc X-ray filament that is probably
made of groups falling onto the main cluster. Interestingly, X-ray
studies based on recent {\it Chandra} and XMM-Newton data (e.g.,
\citealt{kem02,dur05}), provided an evidence of past and present
merger activity for this cluster. 
Our galaxy data showed that only A87 ($\sim30\arcmin$, or 1.9 Mpc 
  from the southwest of the main cluster) is detected as a substructure,
  but some of the galaxies in A87 with $\delta\leq2.0$ were removed for the orbit analysis.
In addition, a small deviation from the scaling relation between X-ray and optical data
  and a small peculiar velocity of the BCG do not imply dynamical non-equilibrium. 
The galaxy orbits are consistent with 
  isotropic orbits within the error at $0.3<r<1.5$ Mpc.
Therefore, it appears that main cluster of A85 is stabilized at present, although it
  experienced a merging in the past, and is not disturbed significantly
  by the current merging event. 
We also find that the determined
  galaxy orbits do not change significantly, if we use different
  X-ray mass profiles as seen in Figure \ref{fig-masscomp}.


\subsection{Abell 779}
A779 is one of the nearest (z$\sim$0.023) clusters and
  shows the faintest X-ray luminosity in our sample.
The velocity dispersion is the smallest ($\sigma_p=491^{+40}_{-36}$ km s$^{-1}$) among our sample, and
  the number of late-types ($N_{gal}=78$) is larger than that of early-types ($N_{gal}=67$).
This cluster has a cD galaxy, NGC 2832, which is known to be at rest in cluster potential \citep{oh01},
  but was not included in the spectroscopic sample of SDSS.
The VDP using all member galaxies with $\delta\leq2.0$ declines at $R\lesssim0.8$ Mpc (see Fig. \ref{fig-disp1}), 
  which is consistent with the results of the previous studies \citep{mah99,rd06}.
The galaxy orbits using all members appear to be consistent with radial or isotropic orbit within the uncertainty
  up to $r\sim r_{200}$, but it is difficult to conclude strongly 
  due to the large error of calculated $\beta_{\rm orb}$.
The orbits of early-types appear to be consistent with being radial or isotropic within the uncertainty
  through the radius,
  but strong conclusion is difficult due to the large error of calculated $\beta_{\rm orb}$.
The orbits of late-types are consistent with being radial in the inner region ($r<0.1$ Mpc) where the observed VDPs
  are not available, 
  but with being isotropic within the error at $r<1$ Mpc and become tangential at outer region.

\subsection{Abell 1650}
A1650 was included in both SDSS and 2dFGRS. 
A recent XMM-Newton observation showed no evidence 
  for spatial temperature variation and surface brightness irregularity, 
  indicating a relaxed system \citep{ty03}. 
Similarly, we found no strong evidence for non-equilibrium (see Table \ref{tab-sum}).
Figure \ref{fig-cmr} shows that ${\rm B_J-R_F}$ color of the BCG
  in this cluster is $\sim0.8$ redder than the color expected from the CMR. 
However, this color deviation of the BCG in this cluster
  is not seen in other studies \citep{md00,pim06}. 
Therefore, the ${\rm B_J-R_F}$ color of the BCG given in 2dFGRS needs further investigation.
The galaxy orbits for all member galaxies are found to be 
  isotropic within the uncertainty or tangential at $r>0.1$ Mpc.
Interestingly, the orbits for early-type and late-type galaxies
  appear to be tangential at $r>0.3$ Mpc.

\subsection{Abell 1651}
A1651 is included in both SDSS and 2dFGRS. 
This cluster is known to be a dynamically relaxed system: 
  regular shape in the {\it ROSAT} PSPC image and 
  symmetric temperature profile from {\it ASCA} data \citep{mar98}. 
Interestingly, a large peculiar velocity ($v_p=350$ km s$^{-1}$) of the BCG
  is found in this study (see Fig. \ref{fig-pec}), 
  indicating a dynamically unrelaxed system. 
\citet{oh01} already identified a peculiar velocity of the BCG ($252-258$ km s$^{-1}$), 
  but they concluded that the peculiar velocity is not large since a significance, 
  $S$ is less than 3. 
They obtained $S\sim1.3$ with 39 member galaxies. 
However, we obtain $S=3.46$ with a large number of member galaxies
  ($N_{gal}=258$) and small velocity error, indicating a significant
  ($S>3$) measured peculiar velocity. 
If we use the galaxies outside the substructure with $\delta\leq2.0$,
  the result change little.
This cluster was included in Las Campanas/Anglo-Australian Telescope Rich Cluster Survey
  (LARCS; \citealt{pim06}), where a recession velocity of the cluster
  and the cD galaxy was determined to be $25466\pm57$ and $25571\pm27$ km s$^{-1}$, respectively. 
However, we obtain the mean velocity of the cluster (biweight location used in this study),
  $v_{\rm cl}=25278\pm95$ km s$^{-1}$ using their data, confirming
  significant peculiar velocity, $v_p=293$ km s$^{-1}$.
In contrast, the galaxy orbits are found to be isotropic 
  within the uncertainty through the radius (see Fig. \ref{fig-orbit1}), 
  which means that the galaxies are in
  equilibrium with the cluster potential, being consistent with the result from X-ray data. 
Therefore, it would be interesting to
  investigate what makes the large peculiar velocity of the BCG.

\subsection{Abell 1795}
A1795 is the most X-ray-luminous cluster in our sample.
A cD galaxy in this cluster was not included in the spectroscopic sample of SDSS.
A large peculiar velocity ($365$ km s$^{-1}$) of the cD galaxy was found
by \citet{hill88}, but later \citet{oh94,oh01} reported the peculiar
velocity is not large ($150-180$ km s$^{-1}$) and its significance
is less than 3. Although detailed {\it Chandra} data showed that
the central core of this cluster is not relaxed (e.g., \citealt{mar01,ett02,fab01}),
other wide-field X-ray data indicate that the cluster is close to
being dynamically relaxed \citep{bh96,bt96,tam01}.
We found that the galaxy orbits derived from all member galaxies with $\delta\leq2.0$ are
  consistent with being tangential at $0.3<r<1.2$ Mpc, 
  but being isotropic within the uncertainty in the outer region.
Interestingly, the orbits of early-types appear to be tangential at $r>0.3$ Mpc,
  while those of late-types to be tangential or isotropic within the error.
The results with different mass profiles are shown in Figure \ref{fig-masscomp}.

\subsection{Abell 1800}
A1800 is one of the member clusters in the Bootes supercluster \citep{ein01},
  and has a relatively small number ratio ($\sim 0.16)$ of late-types to early-types
  among our sample clusters.
There is no strong evidence of dynamically non-equilibrium
  from the substructure tests and the scaling relation between X-ray and optical data.
The galaxy orbits change from radial ($r<0.2$ Mpc),
  to isotropic ($0.2<r<0.8$ Mpc) within the error and to tangential ($0.8<r<1.8$) Mpc.

\subsection{Abell 2034}
A2034 is the most distant cluster (z$\sim0.113$), 
  and has the largest velocity dispersion ($1208_{-76}^{+80}$ km s$^{-1}$)
  in our sample clusters. 
Interestingly, there are only four late-types out of 78 member galaxies.
Since the observed VDP is
  larger than the upper envelope ($\beta_{\rm orb}=0.99$) of the computed one over the radius,
  it is difficult to conclude using top and middle panels.
However, the bottom panel indicates that the galaxy orbits
  are radial in the outer region ($r>1$ Mpc), and those are consistent with isotropic orbits 
  within the uncertainty at $0.2<r<1$ Mpc.

Larger values of observed VDP than the strong radial orbit can be interpreted as non-equilibrium of this cluster,
  although substructure tests and the scaling relation between X-ray
  and optical data in this study imply no strong substructure. 
However, other studies showed direct evidence of dynamical non-equilibrium.
For example, detailed analysis of {\it Chandra} data revealed
  evidence for an ongoing merger \citep{kem03}: northern cold front that is a
  discontinuity of the surface brightness on the northeast edge of
  the cluster, large concentration of galaxies including a cD galaxy
  just ahead of the cold front, and excess of emission to the south of cluster. 
In addition, a cD galaxy in the main cluster is offset
  $\sim1\arcmin$ from the X-ray centroid. 
Unfortunately, the cD galaxy is not included in the spectroscopic sample of SDSS. 
If we adopt a receding velocity ($v_p=33445\pm35$ km s$^{-1}$) of the cD
  galaxy in \citet{mil02}, then the peculiar velocity for this
  galaxy is $477\pm163$ km s$^{-1}$ with the significance of $S=3.378$.
This large peculiar velocity becomes larger if
  we use recession velocity ($v_{\rm cl}=34373$ km s$^{-1}$) of the
  cluster determined in the same reference. 

\subsection{Abell 2199} \label{a2199}
The number of member galaxies ($N_{gal}=754$) in A2199 is the
  largest among our sample clusters, and the number ($N_{gal}=372$)
  of early-types is comparable to that ($N_{gal}=382$) of
  late-types. 
A2199 is known to be one of the richest, regular
  cluster \citep{mar99}, and contains several infalling bound
  subclusters \citep{rines01}. \citet{rines02} derived the VDP
  up to 8 Mpc using $\sim300$ member galaxies in
  A2199, and found that the observed profile is consistent with an
  isotropic orbit based on the caustic mass profile. 
\citet{lok06}, using 180 member galaxies within $R\sim1.1$ Mpc, determined
  $\beta_{\rm orb}=-0.55^{+1.05}_{-2.75}$ based on the joint analysis of
  velocity dispersion and kurtosis, which is consistent with a
  tangential or isotropic orbit. 
We found, using an independent X-ray mass profile and a smoothed velocity profile, that
  the galaxy orbits are isotropic at $0.1<r<0.7$ Mpc within the error,
  and are radial in the outer region.
However, \citet{ben06} found that A2199 has tangentially anisotropic orbit
  in the inner region ($R<1.1$ Mpc), using an X-ray mass profile of
  \citet{mar99} and polynomial fit of VDP.
This discrepancy is demonstrated in Figure \ref{fig-masscomp} (c),
  and is because the mass profile of \citet{mar99} they used
  is larger than that we used \citep{san03} (see \S \ref{discuss} for more detail).
We also investigated the orbital difference between early-type and
  late-type galaxies in A2199. 
Interestingly, both subsamples appear
  to have radial through the radius, indicating they infall from the outer region.

\subsection{Abell 2670}
Abell 2670 was included in both SDSS and 2dFGRS.
We found no strong evidence of substructure from the substructure tests,
  but identified a large peculiar velocity of the BCG, which
  was already found by \citet{bird94a}.
Although X-ray morphology seen by {\it ROSAT} is regular,
indicating a relaxed system (\citealt{hw97}),
\citet{bird94b} concluded that this cluster may consist of
four subclusters that are merging along the line of sight,
using $\sim230$ galaxies in \citet{seg88} catalog.
\citet{hw97} also suggested a merging activity
since the observed VDP is much greater
than that expected from the X-ray mass profile.
In contrast, we found isotropic galaxy orbits, which imply
a relaxed system. The observed dispersion profile in
this study is not significantly different from that in \citet{hw97}.
However, X-ray mass profiles adopted in this study are larger than that
used in \citet{hw97}, thus the value of expected dispersion
is larger than that in \citet{hw97}. Therefore,
other evidence (e.g., irregular gas temperature map)
is needed to confirm the merging activity of this cluster.

\subsection{Abell 2734}
\citet{bur04} found no strong substructure using 125 velocity data 
  from the 2dFGRS data. 
Similar results were found by other authors: \citet{sol99} and \citet{bivet02}
  identified no substructure using 45 and 77 member galaxies in
  ENACS data, respectively. 
However, our 3D substructure test resulted in a
  strong substructure in this cluster.
It is due to the larger spatial coverage than previous studies,
  therefore, the substructure at $R\sim30\arcmin$ ($\sim2$ Mpc) to the south-east
  is included in this study.
Comparison of {\it ROSAT} X-ray image
  with the optical one also shows a good agreement \citep{kol01}.
\citet{dHK96} derived a flat VDP up to R$\sim1.5$ Mpc using 77
  velocity data in ENACS. 
The galaxy orbits appear to be isotropic within the uncertainty 
  through the radius, indicating a dynamical equilibrium, which is consistent with the
  previous studies.

\clearpage

\begin{deluxetable}{ccccccrcrccrccr}
\tabletypesize{\scriptsize} \rotate
\tablewidth{0pc} 
\tablecaption{The Sample of Galaxy Clusters\label{tab-cand}}

\tablehead{
       &             &               &        &                       &       &                 & \multicolumn{2}{c}{All}  && \multicolumn{2}{c}{Early-types} && \multicolumn{2}{c}{Late-types}\\
\cline{8-9} \cline{11-12} \cline{14-15}
       & R.A.        & Decl.         & B-M    &                       & $\overline{cz}$&            & $\sigma_{p}$ & $N_{gal}$ && $\sigma_{p}$ & $N_{gal}$    && $\sigma_{p}$    & $N_{gal}$ \\
Cluster& (J2000)     & (J2000)       & Type   &Survey\tablenotemark{a}& (km s$^{-1}$)  &kpc/arcmin   & (km s$^{-1}$)&           && (km s$^{-1}$)      &        && (km s$^{-1}$)       &          \\
}

\startdata

 A0085 & 00~41~50.09 & $-$09~18~06.8 & I      &S   & $ 16545_{-  64}^{+  65}$ &$  64.3$ & $   926_{-  51}^{+  57}$ &  208 && $   804_{-  54}^{+  61}$ &  170 && $  1114_{- 109}^{+ 108}$ &   38 \\
 A0779 & 09~19~41.28 &   +33~45~46.8 & I-II   &S,2 & $  6947_{-  39}^{+  41}$ &$  28.1$ & $   491_{-  36}^{+  40}$ &  145 && $   471_{-  68}^{+  73}$ &   67 && $   479_{-  40}^{+  47}$ &   78 \\
 A1650 & 12~58~41.09 & $-$01~45~24.8 & I-II   &S,2 & $ 25187_{-  46}^{+  46}$ &$  94.7$ & $   762_{-  39}^{+  39}$ &  258 && $   722_{-  49}^{+  51}$ &  168 && $   829_{-  60}^{+  65}$ &   89 \\
 A1651 & 12~59~21.50 & $-$04~11~40.9 & I-II   &S   & $ 25222_{-  55}^{+  55}$ &$  94.8$ & $   881_{-  44}^{+  46}$ &  258 && $   828_{-  51}^{+  56}$ &  169 && $   996_{-  79}^{+  90}$ &   84 \\
 A1795 & 13~48~52.97 &   +26~35~44.2 & I      &S   & $ 18775_{-  62}^{+  67}$ &$  72.4$ & $   755_{-  39}^{+  43}$ &  163 && $   729_{-  45}^{+  49}$ &  127 && $   806_{-  80}^{+  70}$ &   36 \\
 A1800 & 13~49~21.65 &   +28~06~13.0 & II     &S   & $ 22660_{-  85}^{+  87}$ &$  86.0$ & $   762_{-  45}^{+  49}$ &   93 && $   722_{-  45}^{+  45}$ &   80 && $   786_{- 196}^{+ 184}$ &   13 \\
 A2034 & 15~10~11.74 &   +33~30~52.9 & II-III &S   & $ 33922_{- 156}^{+ 160}$ &$ 123.3$ & $  1208_{-  76}^{+  80}$ &   78 && $  1185_{-  69}^{+  67}$ &   74 && $  1746_{-1006}^{+ 742}$ &    4 \\
 A2199 & 16~28~37.97 &   +39~32~55.3 & I      &S   & $  9221_{-  29}^{+  28}$ &$  36.9$ & $   726_{-  20}^{+  22}$ &  754 && $   703_{-  29}^{+  30}$ &  372 && $   750_{-  30}^{+  31}$ &  382 \\
 A2670 & 23~54~13.39 & $-$10~24~46.1 & I-II   &S,2 & $ 22837_{-  82}^{+  84}$ &$  86.6$ & $   844_{-  49}^{+  51}$ &  106 && $   840_{-  56}^{+  59}$ &   82 && $   870_{- 104}^{+ 100}$ &   24 \\
 A2734 & 00~11~20.71 & $-$28~51~18.0 & III    &2   & $ 18256_{-  76}^{+  81}$ &$  70.5$ & $  1001_{-  51}^{+  52}$ &  192 && $   831_{-  67}^{+  70}$ &  120 && $  1175_{-  79}^{+  82}$ &   68 \\

\enddata
\tablenotetext{a~}{`S' is covered by SDSS and `2' by 2dFGRS.}
\end{deluxetable}

\begin{table}
\centering
\caption{X-ray Properties for Our Sample Clusters\label{tab-xray}}
\begin{tabular}{cccccccc}
\hline\hline
        & $L_{\rm x}$(0.1$-$2.4 keV)\tablenotemark{a} & $\beta$  & $r_c$ & $T_X$ & $T(0)$ & $\gamma$ &      \\
Cluster &  ($10^{44}$ ergs cm$^{-2}$ s$^{-1}$)        &          & (kpc) & (keV) & (keV)  &          &Ref.\tablenotemark{b} \\
\hline


A0085 & ~9.789 & 0.76 & 293.8 &  ... & 10.97 & 1.32 & S03 \\
A0779 & ~0.090 & 0.34 & ~39.4 &  ... & ~3.66 & 1.02 & S03 \\
A1650 & ~7.308 & 0.78 & 214.2 &  ... & ~9.73 & 1.19 & S03 \\
A1651 & ~8.000 & 0.70 & 192.5 &  ... & ~6.87 & 1.10 & S03 \\
A1795 & 10.124 & 0.83 & 288.4 &  ... & ~9.98 & 1.17 & S03 \\
A1800 & ~2.840 & 0.77 & 280.0 & 4.02 &   ... &  ... & R02 \\
A2034 & ~6.850 &  ... &   ... &  ... &   ... &  ... & D03 \\
A2199 & ~4.165 & 0.60 & ~76.9 &  ... & ~4.17 & 1.15 & S03 \\
A2670 & ~1.469 & 0.55 & ~83.8 &  ... & ~5.99 & 1.04 & S03 \\
A2734 & ~2.365 & 0.62 & 151.4 & 3.85 &   ... &  ... & R02 \\

\hline
\end{tabular}
\begin{flushleft}
$^{\mathrm a}$ A0779 \citep{ebe00}, A2034 \citep{ebe98}, A2670 \citep{boh04}, \& other clusters \citep{rb02}. \\
$^{\mathrm b}$ R02 \citep{rb02}, S03 \citep{san03}, and D03 \citep{dem03}.
\end{flushleft}
\end{table}

\begin{deluxetable}{crrrrrrrrrrrrrr}
\tabletypesize{\scriptsize} 
\tablewidth{0pc} 
\tablecaption{Galaxy Number Density Profile for the
Clusters\label{tab-numden}}

\tablehead{
       & $\nu_{0,\rm NFW}$  & $r_{s,\rm NFW}$& $\nu_{0,\rm Her}$  & $r_{s,\rm Her}$&
         $\nu_{0,\rm KEK05}$  & $r_{s,\rm KEK05}$& $\eta_{g,\rm KEK05}$ \\
Cluster& (arcmin$^{-3}$) & (arcmin)& (arcmin$^{-3}$) & (arcmin)& (arcmin$^{-3}$) & (arcmin)& }

\startdata

\multicolumn{8}{c}{All Galaxies} \\
\hline
 A0085 & $ 1.34E-3$ & $  30.2 $ & $ 6.85E-4$ & $  54.6 $ & $ 1.79E-2$ & $  15.6$ & $   9.0 $ \\ 
 A0779 & $ 8.82E-3$ & $   7.4 $ & $ 8.01E-4$ & $  29.1 $ & $ 2.05E-2$ & $  16.0$ & $  12.7 $ \\ 
 A1650 & $ 4.33E-3$ & $  15.3 $ & $ 1.59E-3$ & $  33.6 $ & $ 4.14E-2$ & $  13.0$ & $  10.7 $ \\ 
 A1651 & $ 3.95E-3$ & $  15.3 $ & $ 1.86E-3$ & $  30.1 $ & $ 4.67E-2$ & $  10.4$ & $  10.1 $ \\ 
 A1795 & $ 3.88E-3$ & $  13.7 $ & $ 1.69E-3$ & $  28.4 $ & $ 2.83E-2$ & $  16.4$ & $  11.8 $ \\ 
 A1800 & $ 7.88E-3$ & $   8.0 $ & $ 2.99E-3$ & $  17.6 $ & $ 5.96E-2$ & $   9.0$ & $  11.5 $ \\ 
 A2034 & $ 1.88E-2$ & $   6.3 $ & $ 6.89E-3$ & $  14.0 $ & $ 1.47E-1$ & $   6.3$ & $  11.1 $ \\ 
 A2199 & $ 5.33E-4$ & $  40.5 $ & $ 1.56E-4$ & $  98.1 $ & $ 8.07E-3$ & $  23.3$ & $   9.9 $ \\ 
 A2670 & $ 3.43E-3$ & $  15.6 $ & $ 1.61E-3$ & $  29.7 $ & $ 2.66E-2$ & $  10.9$ & $   9.1 $ \\ 
 A2734 & $ 2.35E-2$ & $   5.6 $ & $ 6.81E-3$ & $  14.4 $ & $ 1.52E-1$ & $   6.6$ & $  11.7 $ \\ 
\hline
\multicolumn{8}{c}{Early-type Galaxies} \\
\hline
 A0779 & $ 7.73E-2$ & $   2.3 $ & $ 1.26E-3$ & $  17.3 $ & $ 1.22E-2$ & $  16.1$ & $  13.6 $ \\ 
 A1650 & $ 9.18E-3$ & $   9.1 $ & $ 2.56E-3$ & $  22.9 $ & $ 3.91E-2$ & $  13.2$ & $  11.7 $ \\ 
 A1795 & $ 3.03E-2$ & $   5.2 $ & $ 5.66E-3$ & $  15.6 $ & $ 4.47E-2$ & $  15.4$ & $  13.4 $ \\ 
 A2199 & $ 5.60E-2$ & $   4.7 $ & $ 2.37E-3$ & $  24.5 $ & $ 3.29E-2$ & $  18.6$ & $  13.2 $ \\ 
\hline
\multicolumn{8}{c}{Late-type Galaxies} \\
\hline
 A0779 & $ 8.96E-6$ & $ 124.4 $ & $ 5.29E-6$ & $ 208.0 $ & $ 2.41E-4$ & $  28.8$ & $   7.4 $ \\ 
 A1650 & $ 4.38E-5$ & $ 119.6 $ & $ 2.54E-5$ & $ 200.1 $ & $ 2.09E-3$ & $  17.0$ & $   6.9 $ \\ 
 A1795 & $ 2.16E-3$ & $   9.9 $ & $ 2.90E-4$ & $  32.9 $ & $ 7.82E-3$ & $  14.8$ & $  11.4 $ \\ 
 A2199 & $ 4.12E-5$ & $ 105.8 $ & $ 2.19E-5$ & $ 189.8 $ & $ 1.20E-3$ & $  26.7$ & $   8.1 $ \\ 

\enddata
\end{deluxetable}

\begin{table}
\centering
\caption{Morphological Parameters of the Clusters\label{tab-mor}}
\begin{tabular}{ccccc}
\hline\hline

Cluster & $\Gamma_A$ & $\Gamma_B$ & $\Theta_2$ & $\epsilon$ \\
        & (kpc)      & (kpc)   & (deg)      &            \\
\hline

 A0085 & $   1358_{-     52}^{+     52}$ & $    868_{-     51}^{+     53}$ & $  154_{-   4}^{+   5}$ & $ 0.36_{- 0.05}^{+ 0.05}$ \\
 A0779 & $   1169_{-     55}^{+     57}$ & $    829_{-     65}^{+     71}$ & $  161_{-   8}^{+   8}$ & $ 0.29_{- 0.07}^{+ 0.06}$ \\
 A1650 & $   2162_{-     81}^{+     85}$ & $   1420_{-     88}^{+     88}$ & $  123_{-   4}^{+   4}$ & $ 0.34_{- 0.05}^{+ 0.05}$ \\
 A1651 & $   2119_{-     80}^{+     75}$ & $   1798_{-     76}^{+     83}$ & $   52_{-  12}^{+  11}$ & $ 0.15_{- 0.05}^{+ 0.05}$ \\
 A1795 & $   1365_{-     65}^{+     69}$ & $   1151_{-     61}^{+     65}$ & $   47_{-  16}^{+  14}$ & $ 0.16_{- 0.06}^{+ 0.06}$ \\
 A1800 & $   1084_{-     54}^{+     57}$ & $    910_{-     59}^{+     63}$ & $   63_{-  17}^{+  18}$ & $ 0.16_{- 0.06}^{+ 0.06}$ \\
 A2034 & $   1237_{-     82}^{+     85}$ & $    902_{-     87}^{+     81}$ & $   23_{-  10}^{+  10}$ & $ 0.27_{- 0.08}^{+ 0.09}$ \\
 A2199 & $   2182_{-     50}^{+     53}$ & $   1646_{-     46}^{+     52}$ & $   20_{-   3}^{+   3}$ & $ 0.25_{- 0.03}^{+ 0.03}$ \\
 A2670 & $    744_{-     41}^{+     41}$ & $    591_{-     37}^{+     39}$ & $   40_{-  10}^{+  13}$ & $ 0.21_{- 0.06}^{+ 0.07}$ \\
 A2734 & $   1413_{-     60}^{+     62}$ & $   1152_{-     57}^{+     61}$ & $   84_{-   9}^{+  11}$ & $ 0.18_{- 0.05}^{+ 0.05}$ \\

\hline
\end{tabular}
\end{table}

\begin{table}
\tiny \centering
\caption{Parameters for Testing the Presence of Substructure\label{tab-sub}}
\begin{tabular}{rrrrrrrrrrrrc}
\hline\hline
Cluster & $I$ & $I_{90}$ & Skewness & clRej & Kurtosis & clRej & AI & clRej & TI & clRej & $\Delta_{obs}$ & f($\Delta_{sim}>\Delta_{obs}$) \\
\hline

 A0085 &  1.03 &  1.03 & $ -0.30$ &   93.0 & $  0.38 $ &  81.4 & $ -0.22 $ &  31.0 &   1.18 &   96.8 &   251 &  0.304 \\
 A0779 &  1.09 &  1.04 & $  0.08$ &   35.6 & $  0.84 $ &  95.2 & $ -1.01 $ &  95.6 &   1.16 &   89.2 &   245 &  0.000 \\
 A1650 &  1.04 &  1.02 & $  0.15$ &   71.4 & $  0.53 $ &  91.2 & $ -0.43 $ &  58.6 &   1.21 &   99.2 &   326 &  0.156 \\
 A1651 &  1.05 &  1.02 & $ -0.31$ &   96.4 & $  0.66 $ &  95.2 & $ -0.61 $ &  77.4 &   0.99 &    7.2 &   345 &  0.075 \\
 A1795 &  0.96 &  1.03 & $ -0.04$ &   16.0 & $ -0.34 $ &  51.4 & $ -0.17 $ &  24.4 &   1.03 &   28.2 &   246 &  0.001 \\
 A1800 &  0.97 &  1.05 & $  0.29$ &   77.2 & $ -0.54 $ &  66.6 & $  0.87 $ &  90.6 &   0.87 &   84.0 &   118 &  0.034 \\
 A2034 &  0.95 &  1.06 & $  0.18$ &   53.0 & $ -0.82 $ &  93.0 & $  0.34 $ &  52.8 &   0.95 &   29.2 &    94 &  0.059 \\
 A2199 &  1.04 &  1.01 & $  0.44$ &   99.8 & $  0.37 $ &  92.2 & $  1.45 $ &  99.2 &   0.93 &   92.6 &  1456 &  0.000 \\
 A2670 &  0.94 &  1.05 & $ -0.01$ &    3.4 & $ -0.66 $ &  86.0 & $ -0.15 $ &  21.0 &   0.94 &   46.6 &   102 &  0.729 \\
 A2734 &  0.97 &  1.03 & $ -0.07$ &   29.4 & $ -0.37 $ &  65.8 & $  0.39 $ &  58.8 &   1.07 &   62.2 &   341 &  0.000 \\

\hline
\end{tabular}
\normalsize
\end{table}

\begin{deluxetable}{ccccccc}
\tabletypesize{\scriptsize} 
\tablewidth{0pc} 
\tablecaption{Summary of Global Kinematic Properties for the Clusters\label{tab-sum}}

\tablehead{
       & \multicolumn{3}{c}{Substructure?}&            & \multicolumn{2}{c}{Equilibrium?}   \\
\cline{2-4} \cline{6-7} 
Cluster& 1D & 2D & 3D                     & Morphology & X-Ray & $v_{pec}$                  }

\startdata

 A0085 &  No   & Yes & No                 & Elongated? & Yes   & Yes  \\
 A0779 &  Yes  & No  & Yes                & Elongated  & No    & ...  \\
 A1650 &  Yes  & Yes & No                 & Elongated  & Yes   & Yes  \\
 A1651 &  Yes  & Yes & Yes                & Spherical  & Yes   & No   \\
 A1795 &  No   & No  & Yes                & Spherical  & No    & ...  \\
 A1800 &  No   & Yes & Yes                & Spherical  & Yes   & ...  \\
 A2034 &  No   & No  & Yes                & Elongated  & Yes   & ...  \\
 A2199 &  Yes  & Yes & Yes                & Elongated  & Yes   & ...  \\
 A2670 &  No   & Yes & No                 & Elongated  & Yes   & No   \\
 A2734 &  No   & Yes & Yes                & Spherical  & Yes   & ...  \\

\enddata
\end{deluxetable}

\clearpage

\clearpage

\clearpage

\clearpage

\end{document}